\documentclass[aps,preprint,floatfix,nofootinbib,showpacs]{revtex4-1}
\pdfoutput=2
\usepackage{graphicx,color}
\usepackage{hyperref}
\usepackage{amsmath}
\usepackage{amsfonts}
\usepackage{amssymb}

\newcommand{\bea}{\begin{eqnarray*}}
\newcommand{\eea}{\end{eqnarray*}}
\begin{document}

\title{$tbW$ Anomalous Couplings in the Two Higgs Doublet Model}
\date{\today}
\author{Abdesslam Arhrib and Adil Jueid}
\affiliation{Universit\'e AbdelMalek Essaadi, 
Facult\'e des Sciences et Techniques, D\'epartement de Math\'ematiques, 
B.P 416 Tangier, Morocco.}
\begin{abstract}
 We make a complete one loop calculation of the $tbW$ couplings 
in the Two Higgs Doublet Model. We evaluate both the anomalous couplings $g_L$
and $g_R$ as well as left handed and right handed component of $tbW$. 
The computation is done in the Feynman gauge 
 using the on-shell scheme renormalization for the Standard Model 
wave functions and parameters. 
We first show that the relative corrections to these 
anomalous couplings are rather small in most regions of the parameter space.
We then analyze the effects of these anomalous couplings on 
 certain observables  such as top quark polarization 
in single top production through $t-$channel as well as $W^\pm$ boson
helicity fractions in top decay. 
\end{abstract}

\maketitle

\section{Introduction} 

Top quark is the heaviest particle discovered by D0 
\cite{Abachi:1995iq} and CDF \cite{Abe:1995hr}
collaborations at the Tevatron-Fermilab with 
mass $m_t=173.21\pm0.51(stat.)\pm0.71(syst.) \text{GeV}$. 
Some of its properties have been studied by the first run of LHC 
 and will get improved by the new LHC run. It is well known that LHC machine
 with  13-14 TeV center of mass energy will act as a top factory 
since the total cross section for top quark pair production will 
reach one nanobarn. 
At the LHC, the top quark production will be two orders of magnitude larger
than in the Tevatron. At low luminosity phase of LHC, one expects about ten
millions top pairs per year and this number will increase during the high 
luminosity phase. Therefore, with such extremely large number of
top anti-top, it is expected that top quark properties (top mass, top spin and
decay rates...) can be examined with very good precision.  

The main decay of the top quark is into W boson and bottom quark. At the
tree level, this decay proceeds  through the left handed V-A charged 
weak interaction which is directly proportional to 
Cabbibo-Kobayachi-Maskawa $V_{tb}$ which can be measured in single top
production. Both in the Standard Model (SM) and 
 Beyond SM, loop effects 
can modify the structure of $tbW$ vertex. Such modifications are typically
described by anomalous couplings $g_L$ and $g_R$ as well as modifications to 
left handed ($V_L$) and  right handed ($V_R$) components of $tbW$.
The QCD corrections to the anomalous coupling $g_R$ have been evaluated 
a while ago in \cite{Li:1990qf}, while the SM electroweak 
and QCD corrections to $g_{L,R}$ and $V_R$  have been  studied in
 \cite{GonzalezSprinberg:2011kx} and \cite{Gonzalez-Sprinberg:2015dea}. 
It turns out that the anomalous couplings $g_L, g_R$ as well as top quark right 
coupling $g_R$ in the $tbW$ are  dominated by the QCD corrections. \\
It is well known that, the anomalous $tbW$ couplings could be probed by 
 measuring the $W$ boson helicity fractions in the top quark decay
\cite{Chatrchyan:2013jna,Aad:2012ky}.
These polarization states are proven to be sensitive to 
new physics effects \cite{Kane:1991bg}. Moreover,
top quark due to its short lifetime, decays before it
hadronizes. Therefore, the information about its polarization may be preserved
in its decay products which can be viewed as top spin analyzers. 

In this study, we are interested in  computing the complete one loop 
contribution to the anomalous $tbW$ couplings 
in the framework of Two Higgs Doublet Model (2HDM). 
We evaluate both the anomalous couplings $g_L$
and $g_R$ as well as left handed $V_L$ and right handed $V_R$ 
component of $tbW$. We stress that, 
evaluation of the top anomalous couplings in the framework of 
the two Higgs Doublet Model (2HDM) has been studied some times ago in 
\cite{Bernreuther:2008us} and recently in \cite{Duarte:2013zfa}.
In \cite{Duarte:2013zfa}, only the computation of the tensorial
anomalous couplings $g_R$ and $g_L$ has been considered. We will
perform, in addition to tensorial couplings $g_{L,R}$, a complete
one-loop computation of the left and right chiral couplings $V_L$ and
$V_R$ and quantify their effects on top quark polarisation in single
production through $t$-channel and $W^\pm$ helicity fractions.

The 2HDM effects are found to be below percent level. 
In the present computation, we perform a comparative study and 
will include all the virtual effect of the 2HDM as well as the real emission 
of photon and gluon in the final state that are necessary for the computation
of the one loop contribution to $V_L$ in order to have infra-red finite result.

There have been several experimental searches for anomalous coupling of the 
top quark. One of the most strongest constraints comes 
from measurement of $Br(\bar{B} \to X_s \gamma)$ \cite{Grzadkowski:2008mf}.
Tevatron also has reported limits on the anomalous couplings in the search 
of new physics in top quark decays \cite{Abazov:2012uga}. We
note also that there are limits from ATLAS and CMS collaborations 
on anomalous couplings from the measurements of the $W^\pm$ helicity fractions
in top quark decay \cite{Chatrchyan:2013jna, Aad:2012ky}. 
In this regard, the first measurement was reported by the CMS collaboration 
\cite{CMS:2012} assuming $V_L=1, g_L=V_R=0$, they have found 
the following value $g_R=0.070\pm0.053(stat.)^{+0.081}_{-0.073}(syst.)$. 
But this measurement suffers from large statistical and systematic 
uncertainties. The sensitivity
of the ATLAS experiment to the anomalous $tbW$ couplings has been 
studied in \cite{AguilarSaavedra:2007rs}.
Finally, we stress here that the anomalous couplings 
might be measured from the measurement of single top production cross section
at the LHC \cite{AguilarSaavedra:2008gt}, from the measurements of
Laboratory frame observables constructed in \cite{Prasath:2014mfa} 
through single top production at the LHC \cite{Jueid:2018wnj}, 
and from the observables that were 
considered for the case of a future $e^- p$ collider \cite{Dutta:2013mva}. \\

In fact, all measurements of top quark properties performed so far
are in perfect agreement with the SM theoretical predictions.
We would like to investigate the top quark $tbW$ anomalous couplings  
 as well as left and right handed $tbW$ couplings in the 2HDM,
 and quantize their effects on some top quark 
observables such as top polarization 
in single top production through $t-$channel as well as $W^\pm$ 
helicity fractions in top decay.

The outline of this paper is the following: In section (\ref{2HDM}), 
we introduce the two Higgs Doublet Model, its parameters
and the constraints that we will use during the numerical 
analysis. In section (\ref{tbW}), 
we describe the experimental status of the anomalous couplings 
and the theoretical set-up 
used in our calculation while in section (\ref{Numerics}), 
we present and discuss our
numerical results. Our conclusions are drawn in section (\ref{Conclusion}).
The appendix is devoted to analytical expression 
for the one-loop anomalous couplings given for the first time 
 in terms of Passariono-Veltman functions 
 and comparison with some results from literature.

\section{The Two-Higgs-Doublet-Model} 
\label{2HDM}
In the Two-Higgs-Doublet Model, two scalar doublets 
under $SU(2)_L$ with Hypercharge $Y_{H_{1,2}}=1/2$
are used to generate fermion and gauge boson masses. 
The inclusion of the two doublets may give rise to sizeable
flavor changing neutral current processes (FCNC) at tree level. 
In order to avoid such tree level FCNC, a discrete symmetry
$Z_2$ (where for example $H_1 \to H_1$ and $H_2 \to -H_2$) is 
imposed \cite{Glashow:1976nt}. Hence, there are $4$ different
combinations of the Yukawa Lagrangian depending on the $Z_2$
charge assignment to the leptons and quarks fields 
\cite{Aoki:2009ha,Branco:2011iw}.  
There are four different models of Yukawa
interactions. In type-I model, only the second doublet
$H_2$ interacts with all the fermions while in type-II model where
the doublet $H_2$ interacts with up-type quarks and $H_1$ interacts
with the charged leptons and down-type quarks.
In type-X model, charged leptons couple to $H_1$ while all the quarks 
couple to $H_2$. Finally, in type-Y model, charged leptons and up-type 
quarks couple to $H_2$ while down-type quarks acquire masses from
their couplings to $H_1$. Given that the Higgs couplings to quarks are the 
 same in type-I (resp type-II) and in type-X (resp type-Y), in
what follow  we will discuss only 2HDM type-I and II.

The Lagrangian representing the Yukawa interactions is given by:
\begin{eqnarray}
 - {\mathcal{L}_{Yuk}} = \bar{q}_L {\mathcal{Y}_{u}} \widetilde{H}_2 u_R + 
\bar{q}_L {\mathcal{Y}_d} H_{d} d_R 
 + \bar{l}_L {\mathcal{Y}_l} H_l l_R + \textrm{H. c.}
 \label{Yukawa}
\end{eqnarray}
Where $H_i, i=l, d$ is either $H_1 \textrm{ or } H_2$ and 
${\mathcal{Y}_i}$ is a set of Yukawa matrices. \\
The most general scalar potential which is gauge-invariant, 
 re-normalizable and CP-invariant is:
\begin{equation} \label{potential}
\begin{split}
V(H_1,H_2) & = \mu_{11}^2 |H_1|^2 + \mu_{22}^2 |H_2|^2 - \mu_{12}^2 (H_1^\dagger H_2 + H_2^\dagger H_1) +  
               \lambda_1 |H_1|^4 + \lambda_2 |H_2|^4 + \lambda_3 |H_1|^2|H_2|^2 + \lambda_4 |H_1^\dagger H_2|^2 \\ 
 & + \frac{\lambda_5}{2} [(H_1^\dagger H_2)^2 + \text{H.c} ]
 \end{split}
\end{equation}
where $\mu_{11,22}^2, \lambda_{i, i=1\ldots4}$ are real parameters while  
$\mu_{12}^2$ and $\lambda_5$ could be complex for CP violating case.
Note that in the above potential the $Z_2$ symmetry is only broken softly by
dimension 2 term $\mu_{12}^2 (H_1^\dagger H_2)$ while dimension four terms are
not introduced in our potential.
The two Higgs doublets $H_1$ and $H_2$ are given by :
\begin{eqnarray}
H_i = \left (\begin{array}{c}
\phi_i^+ \\
v_i + \frac{1}{\sqrt{2}}(h_i + i \omega_i) \\
\end{array} \right)
\qquad , i = 1,2 
\label{Doublet}
\end{eqnarray}
where $v_1 \textrm{ and } v_2$ are the vacuum expectation values of 
the two doublets.
After electroweak symmetry breaking, one has
five additional degrees of freedom; a pair of charged scalar bosons 
$H^\pm$, one CP-odd $A^0$ and two CP-even $h^0, H^0$ where the lightest CP-even 
scalar boson is identified as the SM
Higgs boson. These eigenstates are defined as follow:
\begin{eqnarray}
 \left ( \begin{array}{c}
 h_1 \\
 h_2 \\
 \end{array} \right) = O(\alpha)
 \left ( \begin{array}{c}
 H^0 \\
 h^0 \\
 \end{array} \right), \quad 
 \left ( \begin{array}{c}
 \phi_1^\pm \\
 \phi_2^\pm \\
 \end{array} \right) = O(\beta)
 \left ( \begin{array}{c}
 G^\pm \\
 H^\pm \\
 \end{array} \right), \quad 
 \centering \left ( \begin{array}{c}
 \omega_1 \\
 \omega_2 \\
 \end{array} \right) = O(\beta)
 \left ( \begin{array}{c}
 G^0 \\
 A^0 \\
 \end{array} \right) 
\end{eqnarray}
where: $O(\theta) = \left( \begin{array}{c r}
                             \cos \theta & - \sin \theta \\
                             \sin \theta & \cos \theta \\
                            \end{array} \right) $.\\
The Yukawa Lagrangian in eq. (\ref{Yukawa}) becomes:
\begin{eqnarray}
 - {\mathcal{L}}_{Yuk} = \sum_{\psi=u,d,l} \left(\frac{m_\psi}{v} \kappa_\psi^h \bar{\psi} \psi h^0 + 
 \frac{m_\psi}{v}\kappa_\psi^H \bar{\psi} \psi H^0 
 - i \frac{m_\psi}{v} \kappa_\psi^A \bar{\psi} \gamma_5 \psi A^0 \right) + \nonumber \\
 \left(\frac{V_{ud}}{\sqrt{2} v} \bar{u} (m_u \kappa_u^A P_L +
 m_d \kappa_d^A P_R) d H^+ + \frac{ m_l \kappa_l^A}{\sqrt{2} v} \bar{\nu}_L l_R H^+ + H.c. \right)
 \label{Yukawa-1}
\end{eqnarray}
where $\kappa_i^S$ are the Yukawa couplings in the 2HDM. 
We give, in table (\ref{Yukawa-2}), the values of the couplings in the
four types of Yukawa interactions of the 2HDM with softly broken $Z_2$ symmetry.
\begin{table}
 \begin{center}
  \begin{tabular}{||l|l|l|l|l|l|l|l|l|l||}
   \hline \hline
    & $\kappa_u^h$ & $\kappa_d^h$ & $\kappa_l^h$ & $\kappa_u^H$ & $\kappa_d^H$ & $\kappa_l^H$ & $\kappa_u^A$ & $\kappa_d^A$ & $\kappa_l^A$ \\ \hline
    Type-I & $c_\alpha/s_\beta$ & $c_\alpha/s_\beta$& $c_\alpha/s_\beta$ & $s_\alpha/s_\beta$ & $s_\alpha/s_\beta$ & $s_\alpha/s_\beta$ & $c_\beta/s_\beta$ & 
    $-c_\beta/s_\beta$ & $-c_\beta/s_\beta$ \\ \hline
    Type-II & $c_\alpha/s_\beta$ & $-s_\alpha/c_\beta$& $-s_\alpha/c_\beta$ & $s_\alpha/s_\beta$ & $c_\alpha/c_\beta$ & $c_\alpha/c_\beta$ & $c_\beta/s_\beta$ & 
    $s_\beta/c_\beta$ & $s_\beta/c_\beta$ \\ \hline 
    Type-X & $c_\alpha/s_\beta$ & $c_\alpha/s_\beta$& $-s_\alpha/c_\beta$ & $s_\alpha/s_\beta$ & $s_\alpha/s_\beta$ & $c_\alpha/c_\beta$ & $c_\beta/s_\beta$ & 
    $-c_\beta/s_\beta$ & $s_\beta/c_\beta$ \\ \hline
    Type-Y & $c_\alpha/s_\beta$ & $-s_\alpha/c_\beta$& $c_\alpha/s_\beta$ & $s_\alpha/s_\beta$ & $c_\alpha/c_\beta$ & $s_\alpha/s_\beta$ & $c_\beta/s_\beta$ & 
    $s_\beta/c_\beta$ & $-c_\beta/s_\beta$ \\ \hline \hline
    \end{tabular}
 \end{center}
 \caption{Yukawa couplings in terms of mixing angles in the 2HDM Type I, II, X
   and  Y}
 \label{Yukawa-2}
\end{table}
We will identify the light CP-even Higgs $h^0$ as the 125 GeV SM Higgs, the 
 other parameters of the 2HDM are not yet measured by any experiment, 
 hence we will  apply the following theoretical and experimental 
 constraints on the parameter space of the model:
\begin{itemize}
\item Vacuum stability of the scalar potential \cite{Deshpande:1977rw}.
\item Tree-level perturbative unitarity 
       \cite{Akeroyd:2000wc,Kanemura:1993hm,Kanemura:2015ska}.
\item We will impose constraints on the $\rho$ parameter using the 
PDG update on electroweak fits \cite{Agashe:2014kda}.
\item We impose constraints from the ATLAS measurement \cite{Aad:2015gba} 
 of the signal strength $\mu_{XX}$ defined by :
\begin{eqnarray}
\mu_{XX} = \frac{\sigma (p p \to h^0)^{\text{2HDM}} \Gamma(h^0 \to XX)^{\textrm{2HDM}}}
  {\sigma (p p \to h^0)^{\text{SM}} \Gamma(h^0 \to XX)^{\textrm{SM}}}
 \end{eqnarray}
where $XX$ represents the following channels:  
$W^{\pm*} W^\mp, ZZ^*, \gamma \gamma, \textrm{ and } \tau^+ \tau^-$. 
 While $\sigma(p p \to h^0)$ 
includes the following Higgs production mechanisms
at the LHC: $ggF$, Vector Boson fusion VBF, 
Higgs-strahlung $W^\pm h^0,Z h^0 $  and  $t\bar{t} h^0$.
 \item We will use the results of indirect constraints on the charged 
 Higgs boson 
 mass from processes at the one-loop order, e.g 
 $b\to s \gamma \textrm{ and } R_b$ \cite{Baak:2011ze,Hermann:2012fc,
 Misiak:2006zs, Mahmoudi:2009zx, Freitas:2012sy, Denner:1991ie,Haber:1999zh, Misiak:2015xwa}.
In our analysis, we assume that $m_{H\pm}\geq 480$ GeV in 2HDM type-II.

 \item Constraints from direct searches of charged Higgs bosons at 
LEP \cite{Abbiendi:2013hk} and the LHC 
 \cite{TheATLAScollaboration:2013wia, Aad:2012tj, Chatrchyan:2012vca} 
will be used.
\end{itemize}

\section{Anomalous $tbW$ couplings} 
\label{tbW}
Owing to Lorentz invariance, the amplitude of top quark decay 
$t(p_t) \to b(p_b) W^+(q)$ can be written as:
\begin{eqnarray}
 \mathcal{M}(t\to bW^+) = \frac{-e}{\sqrt{2} \sin \theta_W} 
\bar{u}_b(p_b) \left[(V_L P_L + V_R P_R)\gamma^\mu
 + \frac{i \sigma^{\mu\nu} q_\nu}{M_W} (g_L P_L + g_R P_R) \right] 
u_t(p_t) \epsilon_\mu^*(q),
\label{anomalous}
 \end{eqnarray}
where $P_{R,L}=\frac{1}{2} (1\pm\gamma_5)$ are the projection
operators, $p_t, p_b \textrm{ and } q = p_t-p_b$ 
are respectively the four-momenta of the top, bottom and $W^+$ boson.
The three particles are assumed to be on-shell. For the case of 
$W^+$ being off shell, there are two-additional terms which should be added
to the matrix elements\footnote{In other words, two
form factors $f_L$ and $f_R$ have to be added as follows 
$\bar{u}_b(p_b) \frac{i \sigma^{\mu\nu} (p_t + p_b)_\nu}{M_W} 
(f_L P_L + f_R P_R) u_t(p_t) \epsilon_\mu^*(q)$}
in eq. (\ref{anomalous}).
At tree level, in the SM, $V_L = V_{tb}$ and $V_R=g_R=g_L=0$,
 while radiative corrections in the SM induce non-zero values for 
$V_R, g_R \textrm{ and } g_L$. Note that renormalizable theories 
beyond the SM might induce non-zero values for the right chiral 
coupling $V_R$ even at tree level, but $g_R \textrm{ and } g_L$ 
 have to be induced only at one-loop order.
 Before discussing the details of our calculations,
we recapitulate the experimental status of the direct searches
for the anomalous couplings as well as the indirect 
constraints coming from one-loop induced processes. \\
One of the most strongest constraints comes 
from $Br(\bar{B} \to X_s \gamma)$ \cite{Grzadkowski:2008mf}.
The enhancement factor $m_t/m_b$ implies that 
these constraints are stronger 
for $V_R \textrm{ and } g_L$ and rather weaker for $g_R$:
\begin{eqnarray} 
&& -0.15 \leq \textrm{Re}(g_R) \leq 0.57 \nonumber\\
&& -7 \times 10^{-4} \leq V_R \leq 2.5 \times 10^{-3} \nonumber\\
&& -1.3 \times 10^{-3} \leq g_L \leq 4 \times 10^{-4} 
\end{eqnarray} 
The above limit on $V_R$ and $g_L$ would be improved if one can measure 
more accurately  
$Br(\bar{B} \to X_s \gamma)$ at the Super B factory \cite{Akeroyd:2004mj} 
and/or LHCb.\\
There are also $2\sigma$ limits
available for $g_{L,R}$ using LHC simulations \cite{AguilarSaavedra:2006fy}; 
 $$-0.026 < g_R < 0.024 \quad  \text{and} \quad -0.058 < g_L < 0.026$$

In the search  of new physics in top quark decays \cite{Abazov:2012uga}, 
Tevatron has reported $95\% \text{ CL}$ limit on anomalous couplings as follow:
 \begin{eqnarray} 
&& |V_R|^2 < 0.30 \qquad , \quad |g_L|^2 < 0.05 \quad \text{and} \quad 
|g_R|^2 < 0.12
\end{eqnarray}
It was assumed $V_L=V_{tb}$.

There are also $95\%$ limits \cite{Fabbrichesi:2014wva} 
on all the anomalous couplings including the 
left chiral coupling $V_L$ from a global fit to the experimental data 
which corresponds to single top production cross section (all the channels
were included) and $W^\pm$ helicity fraction at Tevatron and the LHC. These
limits are:
\begin{eqnarray}
&& -0.142 \leq g_R \leq 0.023 \quad , \quad -0.081 \leq g_L \leq 0.049, \nonumber\\
&& 0.902 \leq V_L \leq 1.081 \quad \text{ and } \quad -0.112 \leq V_R \leq 0.162
\end{eqnarray}
Moreover, global fit of the anomalous $Wtb$ couplings has
been performed in \cite{Cao:2015doa} where correlations among the different 
effective operators have been investigated. \\

On the other hand, limits on tensorial anomalous
couplings $g_L$ and $g_R$ have been studied in 
\cite{Cirigliano:2016njn, Cirigliano:2016nyn}
by combining several constraints from $b\to s\gamma$, 
helicity fractions, single  top production, 
electroweak precision test (mainly from the S-parameter) 
and the electric dipole moments. 
It was found that the real part of $g_R$ is strongly constrained
by the helicity fractions while the strongest 
constraint on $\text{Re}[g_L]$ comes from 
$b\to s \gamma$ branching ratio. On the 
 other hand, the imaginary part of tensorial couplings is 
severely constrained
by the electric dipole moments (EDM); 
e.g the strongest constraint on $\text{Im}[g_L]$
comes from neutron EDM while on $\text{Im}[g_R]$ 
comes from electron EDM.

The ATLAS collaboration \cite{Aad:2015yem} has reported  
$95\%$ CL limits on the ratios of the anomalous couplings $g_R$ and 
$V_L$ from the measurement
of the double differential decay rate of the top quark in single
 top production through
$t$-channel process at $\sqrt{s} = 7 \text{ TeV }$ taking $V_R=g_L=0$. 
The limits are :
\begin{eqnarray}
\text{Re}\left[\frac{g_R}{V_L}\right] \in [-0.36,0.10] \text{ and } 
\text{Im}\left[\frac{g_R}{V_L}\right] \in [-0.17,0.23]
\end{eqnarray}
Recently, Ref.~\cite{Birman:2016jhg} puts $95 \%$ CL limits on the real 
and imaginary part 
of the anomalous couplings which were obtained from a global 
fit to data using the following observables: 
\begin{itemize}
\item $t$-channel single top production cross section at the LHC at $\sqrt{s}
  = 7, 8 \textrm{ and } 13$ TeV 
and at Tevatron   $\sqrt{s} = 1.96$ TeV, 
\item $s$-channel $tW$ associated production at both the LHC $7 \oplus 8$ TeV
  and Tevatron
\item Results from $W^\pm$ helicity fractions in $t\bar{t}$ production at $\sqrt{s} = 8$ TeV 
\item Expected results corresponding the $t$-channel production cross section
at $\sqrt{s} = 14 \text{ and } 33$ TeV assuming that $V_L = V_{tb} \simeq 1$. 
\end{itemize}
We stress that these limits are rather 
weak for $V_R, \textrm{ and } g_L$ and  slightly stronger 
for the case of $g_R$.

\begin{figure}[!h]
 \centering
 \includegraphics[width=14cm, height=12cm]{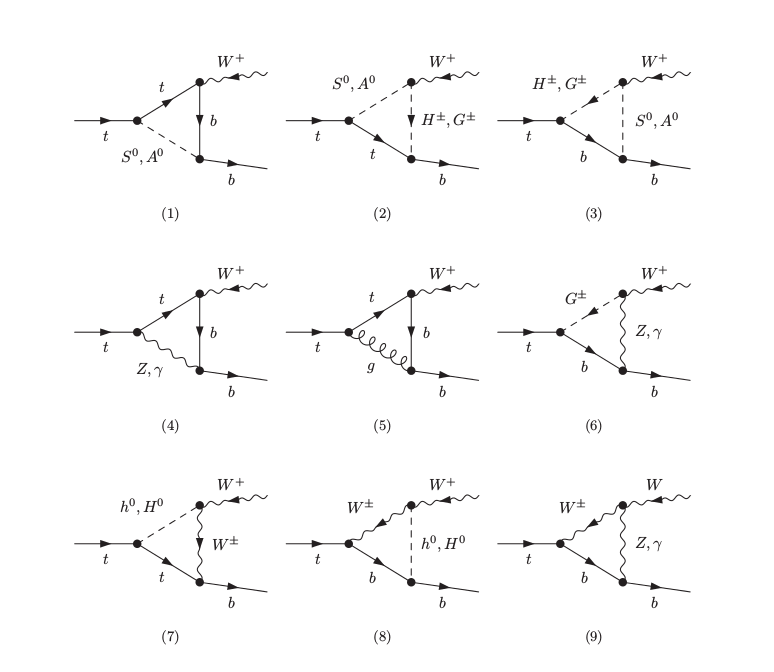}
 \caption{Feynman diagrams that contribute to one loop $tbW$ coupling
 in the 2HDM}
 \label{diagrams}
\end{figure}

We will do a complete analysis of all the anomalous couplings 
present in eq. (\ref{anomalous}). The corresponding Feynman 
 diagrams are depicted in figure~\ref{diagrams}.
For the calculation of $g_L, g_R \textrm{ and } V_R$, 
there is no need to renormalize the theory since these couplings are
absent at tree level. In fact,  infrared divergences are 
also absent in the case of these couplings.
For instance, in the case of $g_L$ and $g_R$, contributions from diagrams
with exchange $t W^\pm \gamma$ and $t G^\pm \gamma$ 
(from $b W^\pm \gamma$ and $b G^\pm \gamma$)  are individually
infrared divergent but their sum
is infrared finite. On the other hand, in the case of $V_R$, all the 
diagrams involving the photon/gluon are IR finite. \\
Before computing  the anomalous couplings 
in the framework of the 2HDM, we have calculated them
in the SM and compared with the results of \cite{GonzalezSprinberg:2011kx} for 
the case of $g_L$ and $g_R$ and with \cite{Gonzalez-Sprinberg:2015dea} 
for the case of the right chiral coupling 
$V_R$. The numerical values of $g_L$, $g_R$ and $V_R$ in the 
SM are tabulated in appendix (\ref{appen-1}) while 
their analytical expressions are given for the first time in terms of
Passarino-Veltamn functions in appendix (\ref{appen-2}). \\

The one-loop corrections to $V_L$ involves 
divergent integrals. In order to get meaningful results
we should add appropriate counter-term to the bare coupling. 
In order to achieve that, we will be working in the on-shell 
renormalization scheme \cite{onshell1, onshell2} where necessary redefinition
of the fields and parameters is performed such that the total  amplitude 
(un-renormalized and counter term) is UV-finite. The counter-term of $tbW$
coupling is given by:
\begin{eqnarray}
 \delta \mathcal{M}_{tbW} = \bar{u}_t(p_t) i e \gamma^\mu P_L \delta C_- u_b(p_b) \epsilon_\mu^*(q)
 \label{counter}
\end{eqnarray}
where $\delta C_-$ is:
\begin{eqnarray}
 \delta C_- = \frac{1}{\sqrt{2} s_W} \left(\delta Z_e - \frac{\delta s_W}{s_W} + \frac{1}{2} \delta Z_W + \frac{1}{2} (\delta Z^{t,L\dagger} + \delta Z^{b,L})\right)
\end{eqnarray}
where we have assumed that $V_{tb}=1$ and  $\delta V_{tb}=0$. 
The renormalization constants 
$\delta Z_e, \delta s_W, \delta Z_W, \delta Z^{t,L\dagger} 
\textrm{ and } \delta Z^{b,L}$ are determined as usual by 
suitable mass and field renormalization conditions. \\
The Feynman diagrams and the corresponding amplitudes have been 
generated with FeynArts and FormCalc packages \cite{FA2}. 
The output was passed to LoopTools \cite{FF} for 
numerical integration of the one-loop functions. 
UV divergences and renormalization scale independence
have been checked analytically with FormCalc and 
numerically with LoopTools. However, 
due to the contribution of virtual photons and gluons, 
the corrections to $V_L$ are infrared divergent. These IR divergences
 are canceled after introducing real photons and gluons emissions 
in the final state.
We have checked that indeed, the total amplitude consisting of virtual, 
 soft and hard photons/gluons emissions are independent of the effective 
cutoff $\lambda_{IR}^2$. This cancellation has been checked analytically 
by computing  the IR divergent part of the three-points Passarino Veltman 
function $C_0$ in  the soft limit using analytical expressions 
from \cite{Dittmaier:2003bc} and the real 
(soft and hard) emission factors extracted 
from \cite{onshell2}. With LoopTools, We checked numerically that the total contribution:
\begin{eqnarray}
 2 \text{Re} (\mathcal{M}_{\text{tree}}^* \mathcal{M}_{\text{virtual}}) + |\mathcal{M}_{\text{real}}|^2
 \label{IR-check}
\end{eqnarray}
is independent of $\lambda_{IR}^2$ by computing the sum (\ref{IR-check}) for different values of 
$\lambda_{IR}^2 \in [10^{-10}:10^6]$ and have
found that the sum is $\lambda_{IR}^2$ independent.
\section{Numerical Results} 
\label{Numerics}
The input parameters of the SM are taken from 
the Particle Data Group \cite{Agashe:2014kda}:
\begin{center}
\begin{tabular}{l|l}
 \hline \hline 
  $m_t = 173.21$ GeV & $m_b = 4.66$ GeV \\ \hline
  $M_W = 80.385$ GeV & $M_Z = 91.1876$ GeV \\ \hline
  $\alpha_S = 0.118$ & $m_H = 125$ GeV \\
  \hline \hline
 \end{tabular} \\
 \end{center}
while the parameter space of the 2HDM 
is scanned over the range specified in table \ref{range-1}.
\begin{table}[!h]
 \begin{center}
  \begin{tabular}{||l|l||}
   \hline
   Type-I & Type-II \\ \hline
   $100 \textrm{ GeV } \leq m_{H^\pm} \leq 900 \textrm{ GeV }$ & $480 \textrm{ GeV } \leq m_{H^\pm} \leq 900 \textrm{ GeV }$ \\ \hline
   $90 \textrm{ GeV } \leq m_{A^0} \leq 900 \textrm{ GeV }$ &  $90 \textrm{ GeV } \leq m_{A^0} \leq 800 \textrm{ GeV }$ \\ \hline
   $125 \textrm{ GeV } \leq m_{H^0} \leq 900 \textrm{ GeV }$ & $125 \textrm{ GeV } \leq m_{H^0} \leq 900 \textrm{ GeV }$  \\ \hline
   $0 \leq \sin ({\beta-\alpha}) \leq 1$ &  $0 \leq \sin ({\beta-\alpha}) \leq 1$ \\ \hline
   $1 \leq \tan \beta \leq 30$ & $1 \leq \tan \beta \leq 30$  \\ \hline
  $-25 \leq \lambda_5 \leq 25$ &  $-25 \leq \lambda_5 \leq 25$ \\ \hline
  \end{tabular}
 \end{center}
 \caption{Parameter space of the Two-Higgs-Doublet model Type-I and -II over which the scan has been performed}
 \label{range-1}
\end{table}

For our numerical analysis, we define the following ratios 
$\Delta \mathcal{O}_i$:
\begin{eqnarray}
 \Delta \mathcal{O}_i = \frac{\mathcal{O}_i^{2HDM}-\mathcal{O}_i^{SM}}{\mathcal{O}_i^{SM}}
\end{eqnarray}
where $\mathcal{O}_i = 
\text{Re}(g_L), \text{Re}(g_R), \text{Re}(V_R)
\textrm{ and } \text{Re}(V_L) + V_{tb}$.

\begin{figure}[tbp]
\centering 
\includegraphics[width=.32\textwidth]{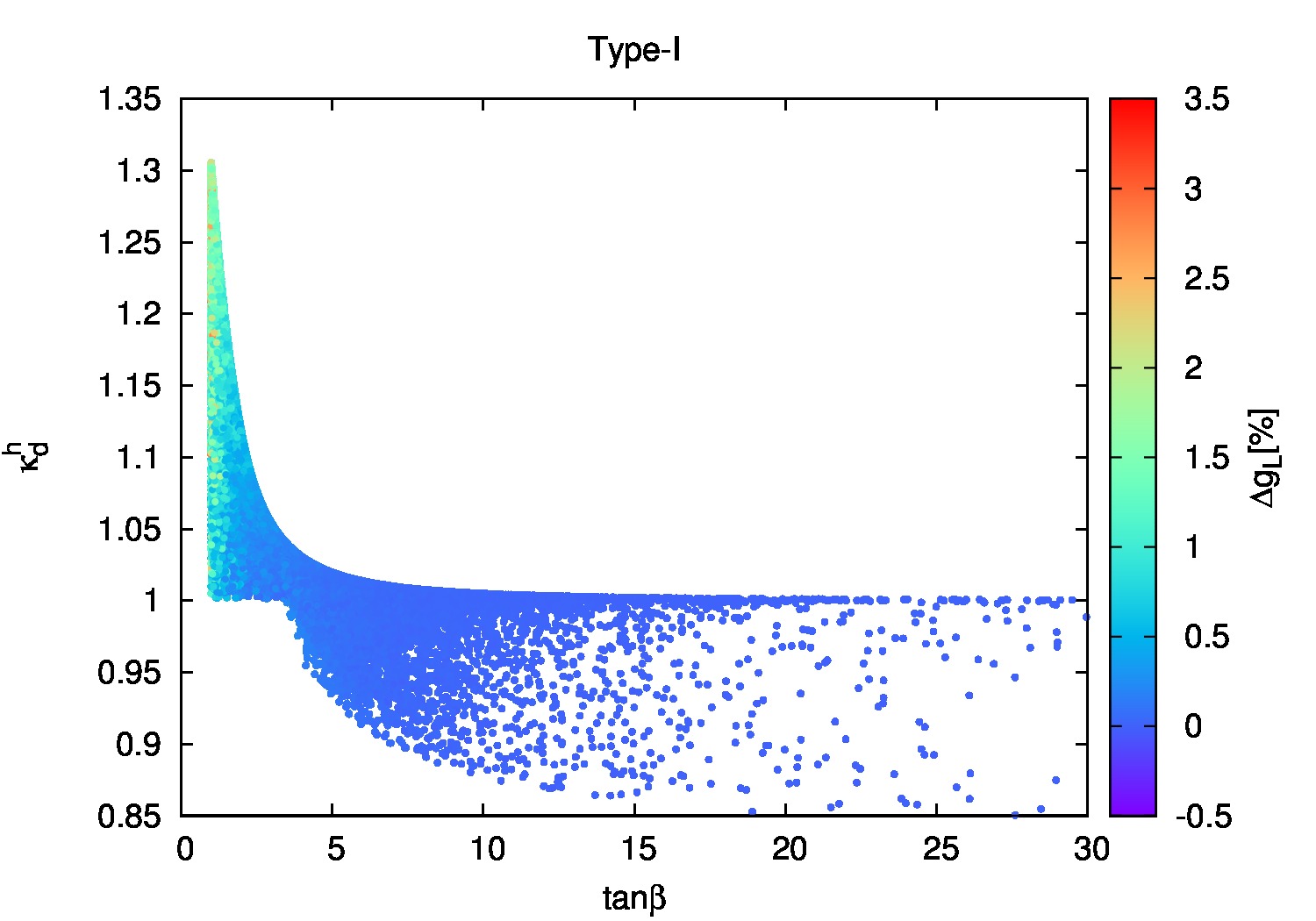}
\hfill
\includegraphics[width=.32\textwidth]{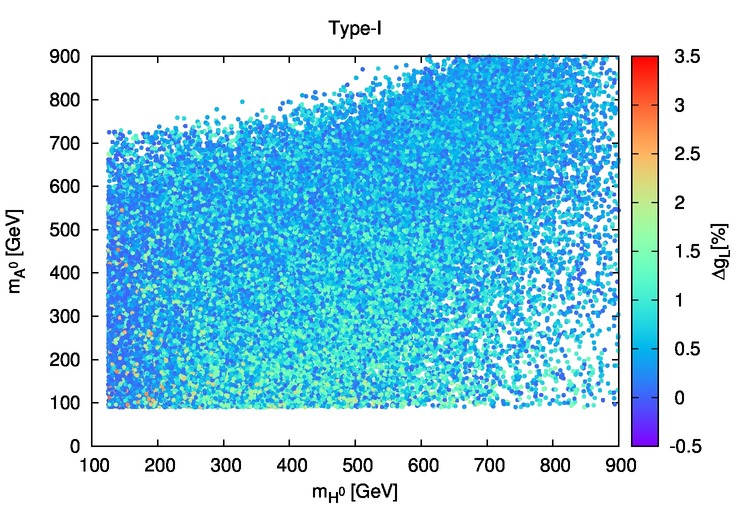}
\hfill
\includegraphics[width=.32\textwidth]{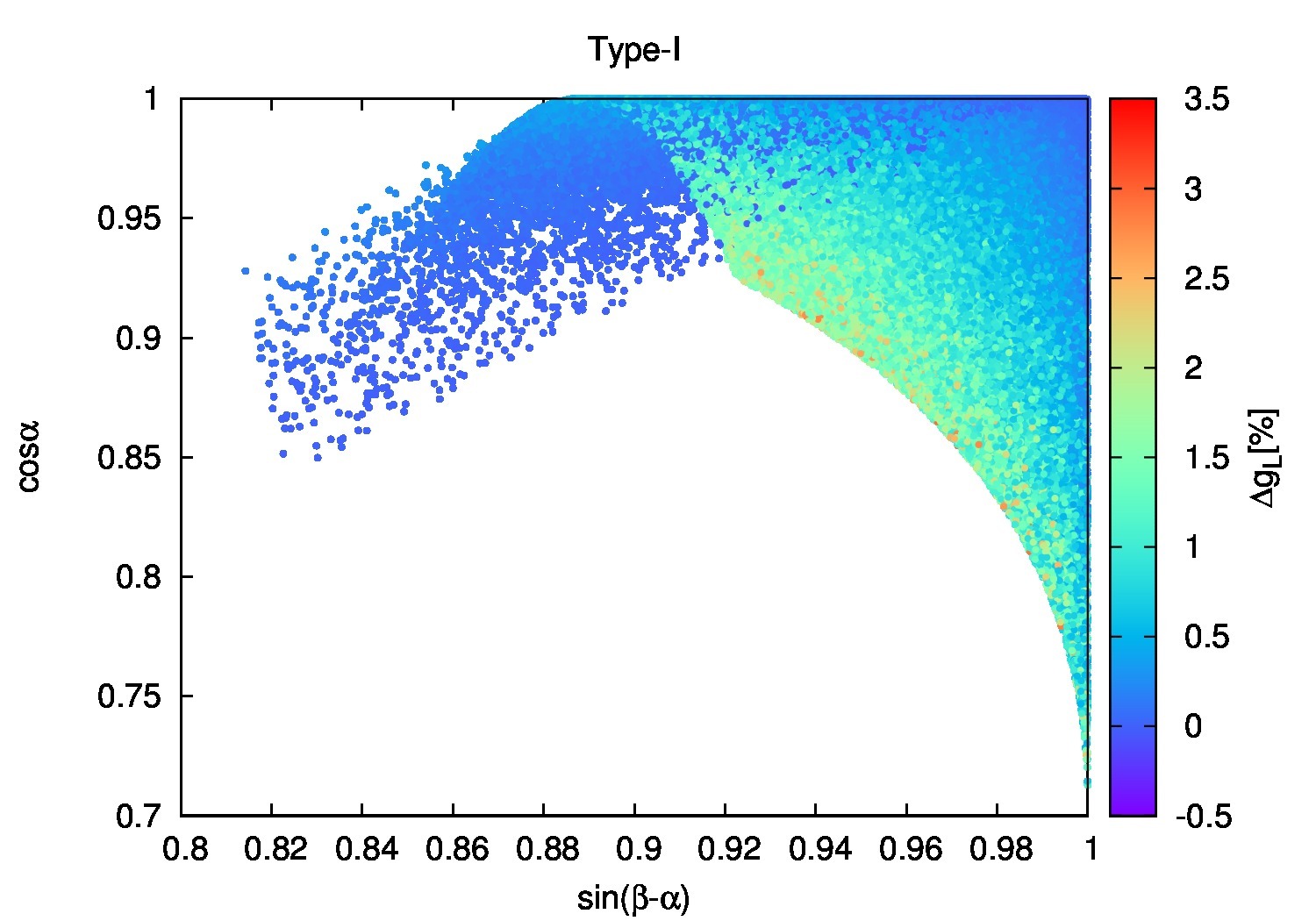}
\hfill
\includegraphics[width=.32\textwidth]{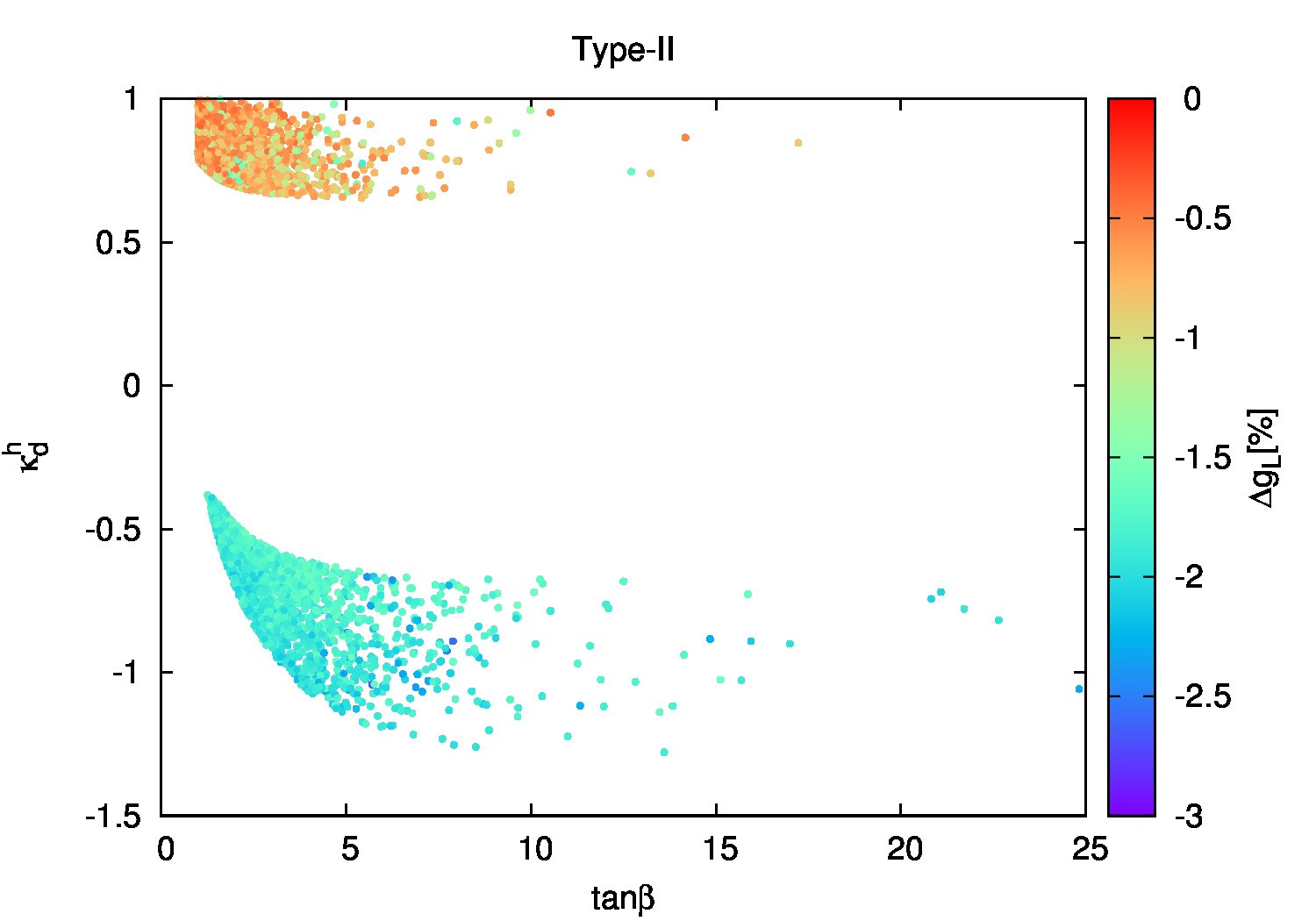}
\hfill
\includegraphics[width=.32\textwidth]{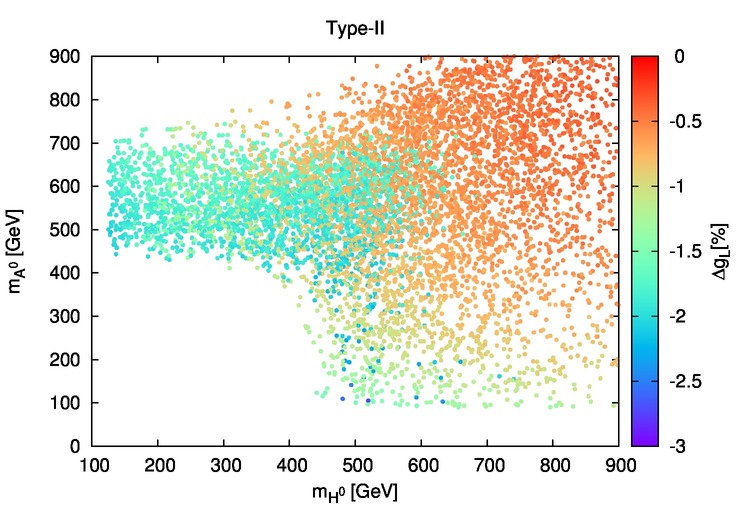}
\hfill
\includegraphics[width=.32\textwidth]{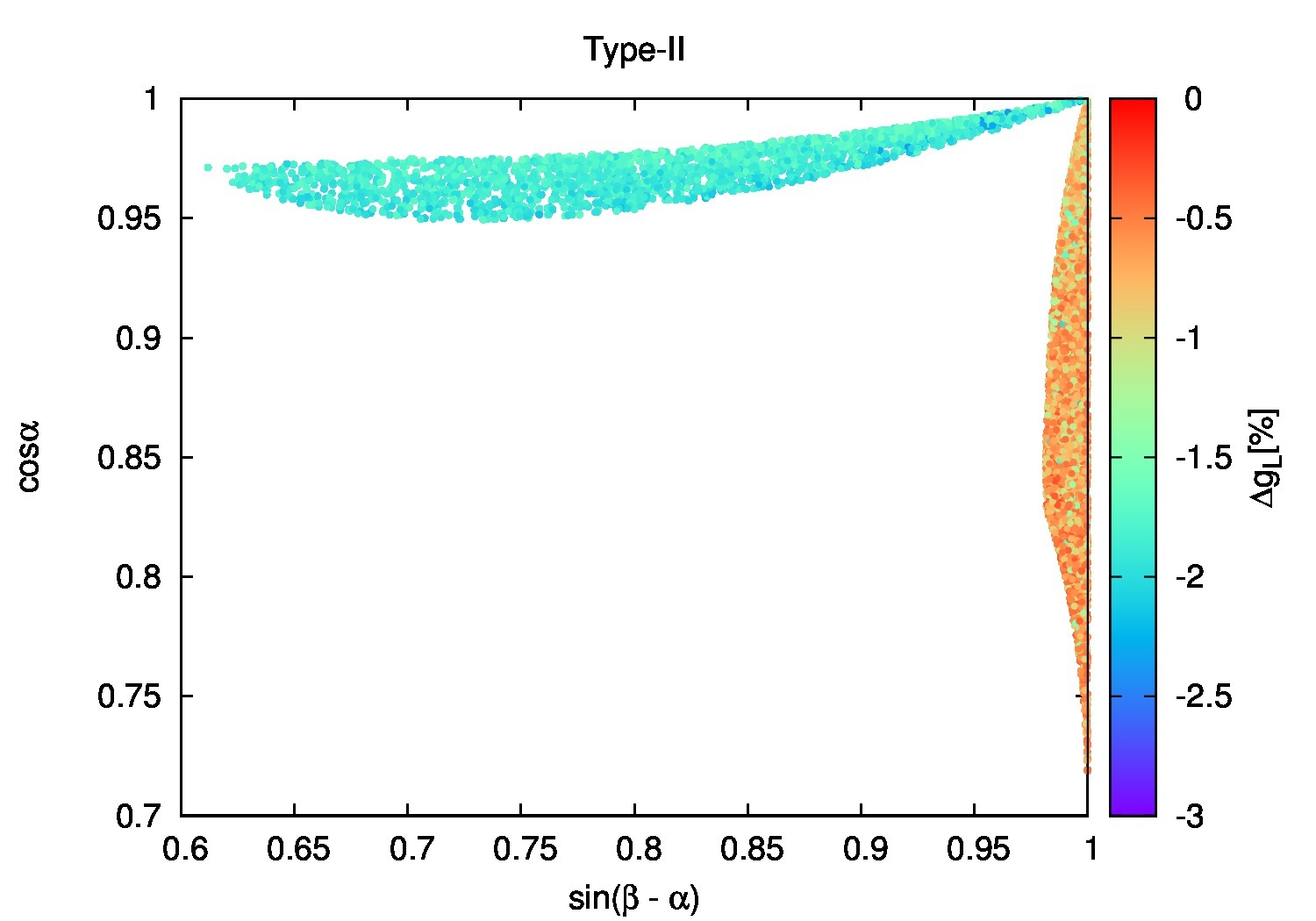}
\caption{\label{gL-THDM} Relative contribution $\Delta g_L$ in 
2HDM type-I (upper panels)
and type-II (lower panels)
shown as a scatter plot in the ($t_\beta$, $\kappa_d^h$) plan (left), 
($m_{H^0},m_{A^0}$) plan (middle) and ($s_{\beta-\alpha}, c_\alpha$) plan (right)}
\end{figure}
\subsection{Anomalous couplings}

The relative correction $\Delta g_L$  is shown in
figure~\ref{gL-THDM} in different plan ($\tan\beta$, $\kappa_d^h$) 
(left panels),
($m_H,m_A$) (middle panels)  and ($\sin (\beta-\alpha), \cos\alpha$). 
Upper panels are for 2HDM-I and lower panels for 2HDM-II. 
One can see that $g_L$ gets enhancement for type-I 
(in most regions of the parameter space) 
while it is always suppressed with respect to the SM for 2HDM type-II. 
Most of the regions in the parameter space correspond to 
$\Delta g_L < 2\%$.
In type-I, $\Delta g_L$ reaches $3.5 \%$ for $\tan \beta \sim 1$ 
(left panel) and for all values of $\kappa_d^h$ between $1 \text{ and } 1.3$ and
also for $m_{H^0,A^0} \in [100:300] \text{ GeV}$ 
(middle panel).
A decoupling behavior is easily observed for large values of
$\tan \beta\geq 25$, $\kappa_d^h \sim 1$ and
also for heavy scalars $m_{A^0,H^0} > 700 \text{ GeV}$.\\ 
In 2HDM type-II, We see that $\Delta g_L$ is always negative while it 
approaches $0\%$, (SM regime), for $\kappa_d^h\sim 1$ and $\tan \beta
\sim 1$. \\
\begin{figure}[tbp]
\centering 
\includegraphics[width=.32\textwidth]{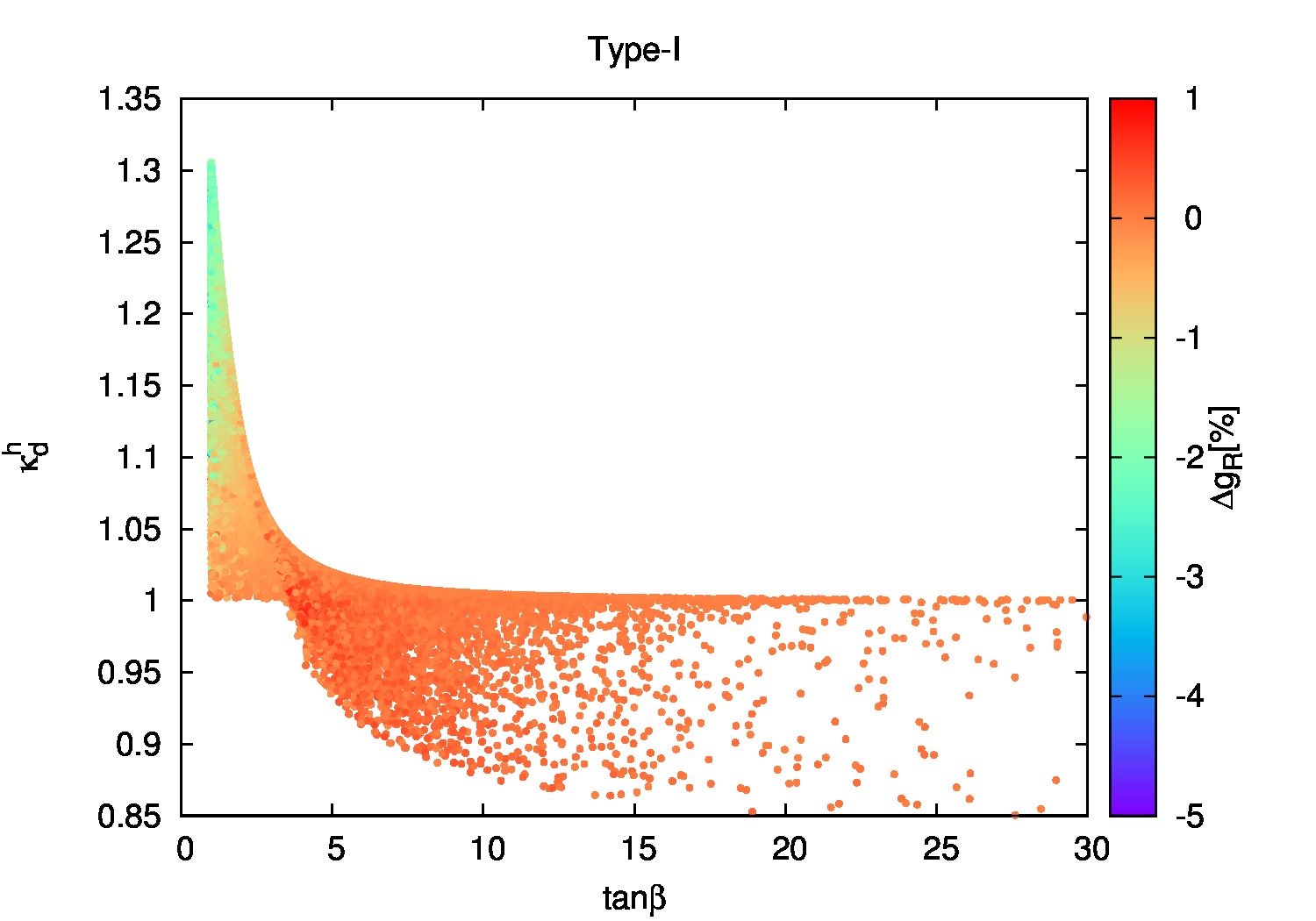}
\hfill
\includegraphics[width=.32\textwidth]{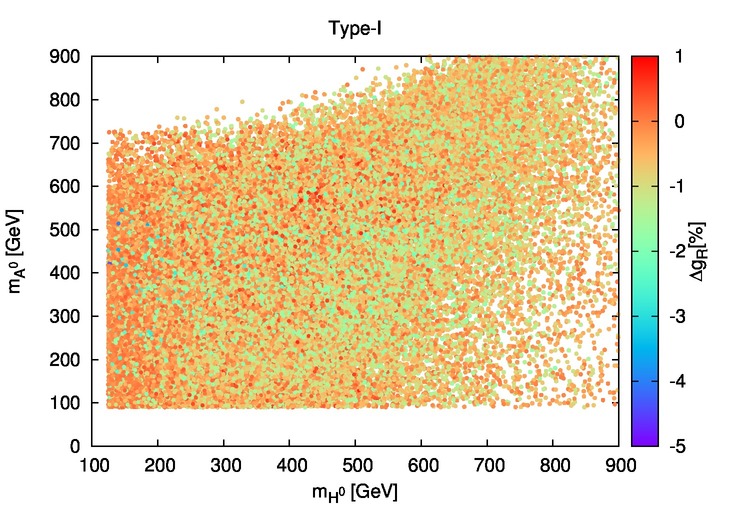}
\hfill
\includegraphics[width=.32\textwidth]{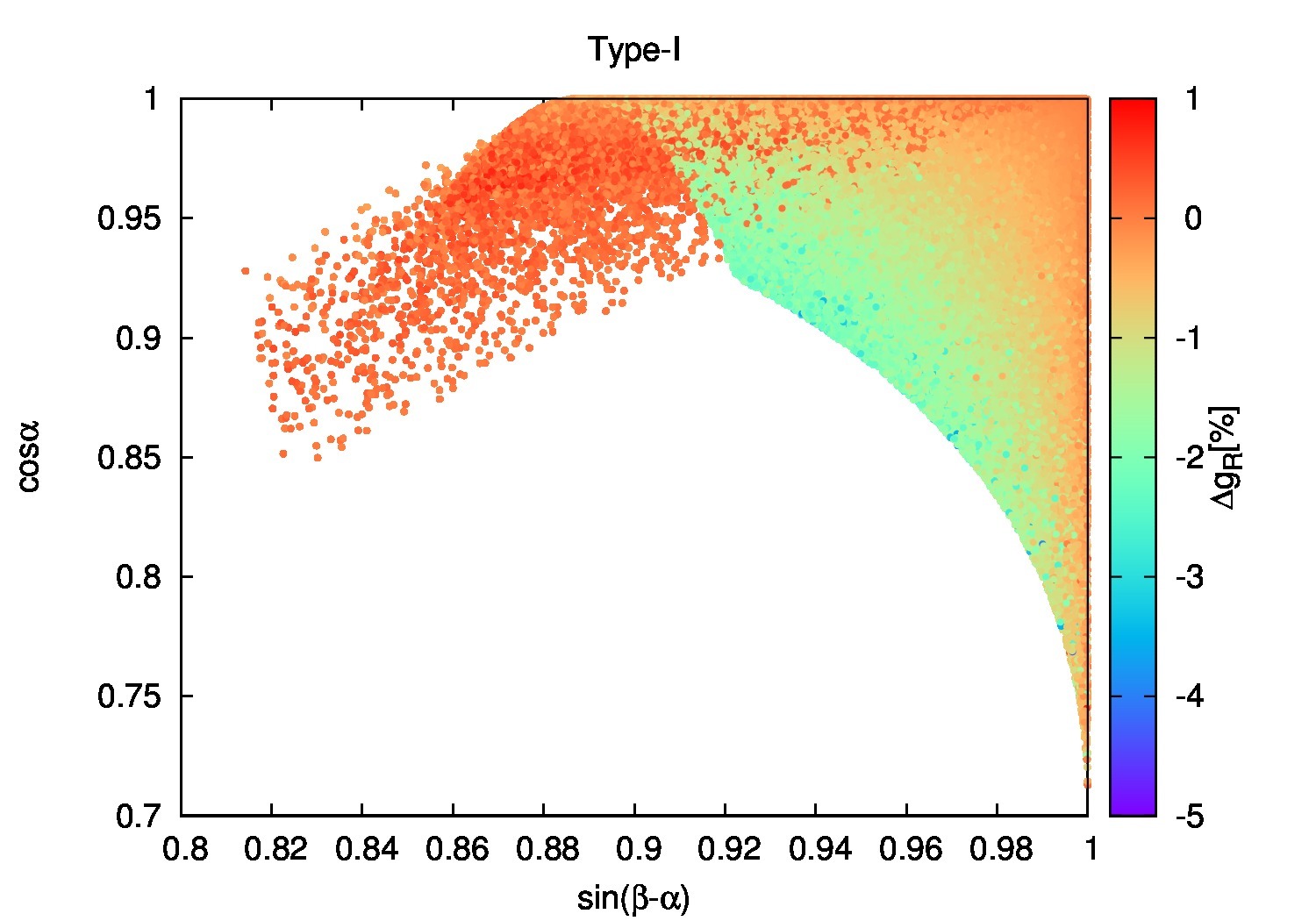}
\hfill
\includegraphics[width=.32\textwidth]{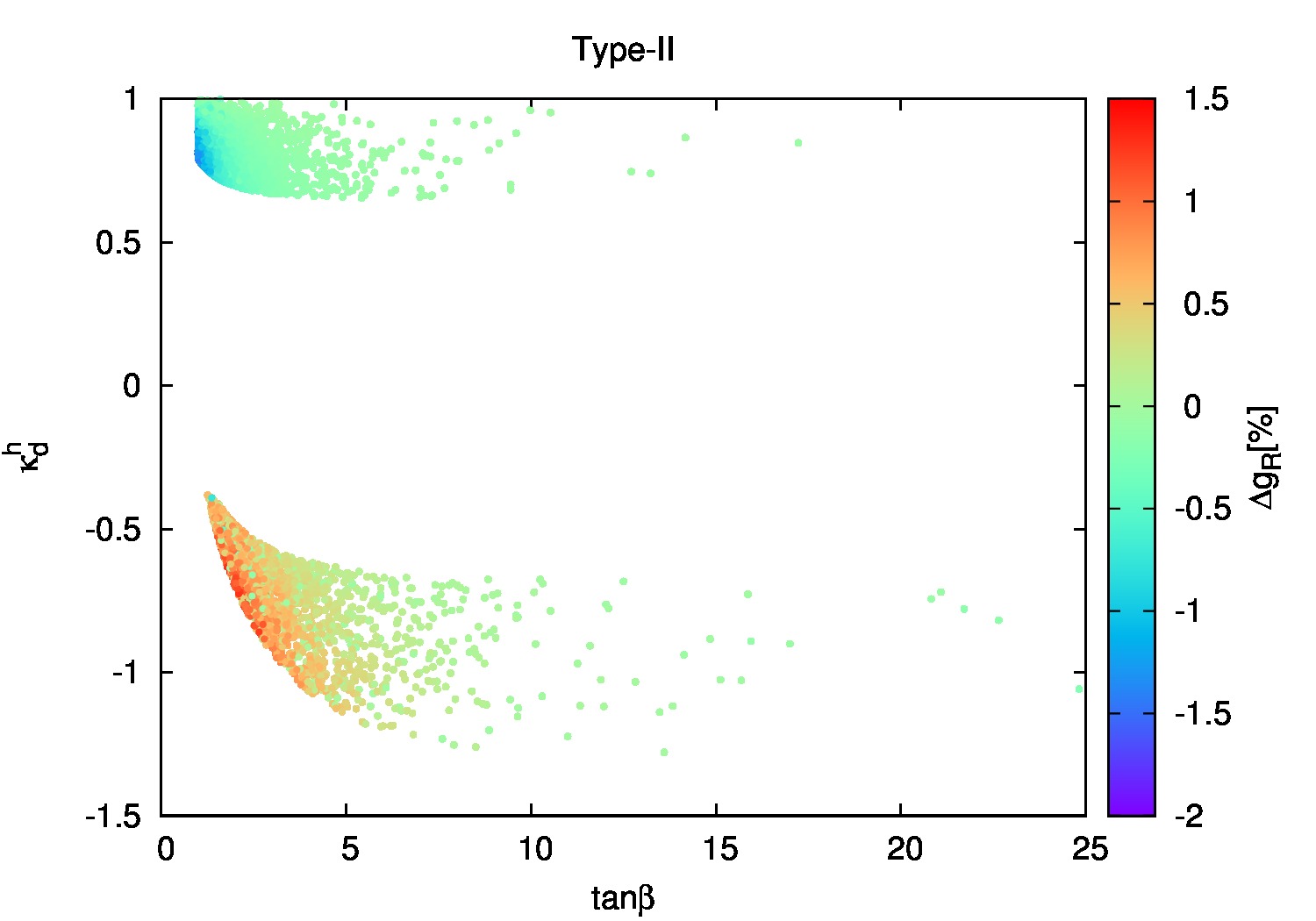}
\hfill
\includegraphics[width=.32\textwidth]{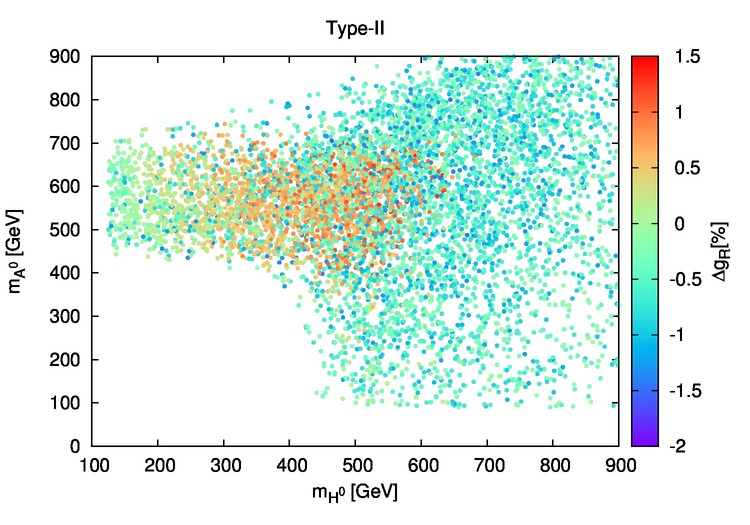}
\hfill
\includegraphics[width=.32\textwidth]{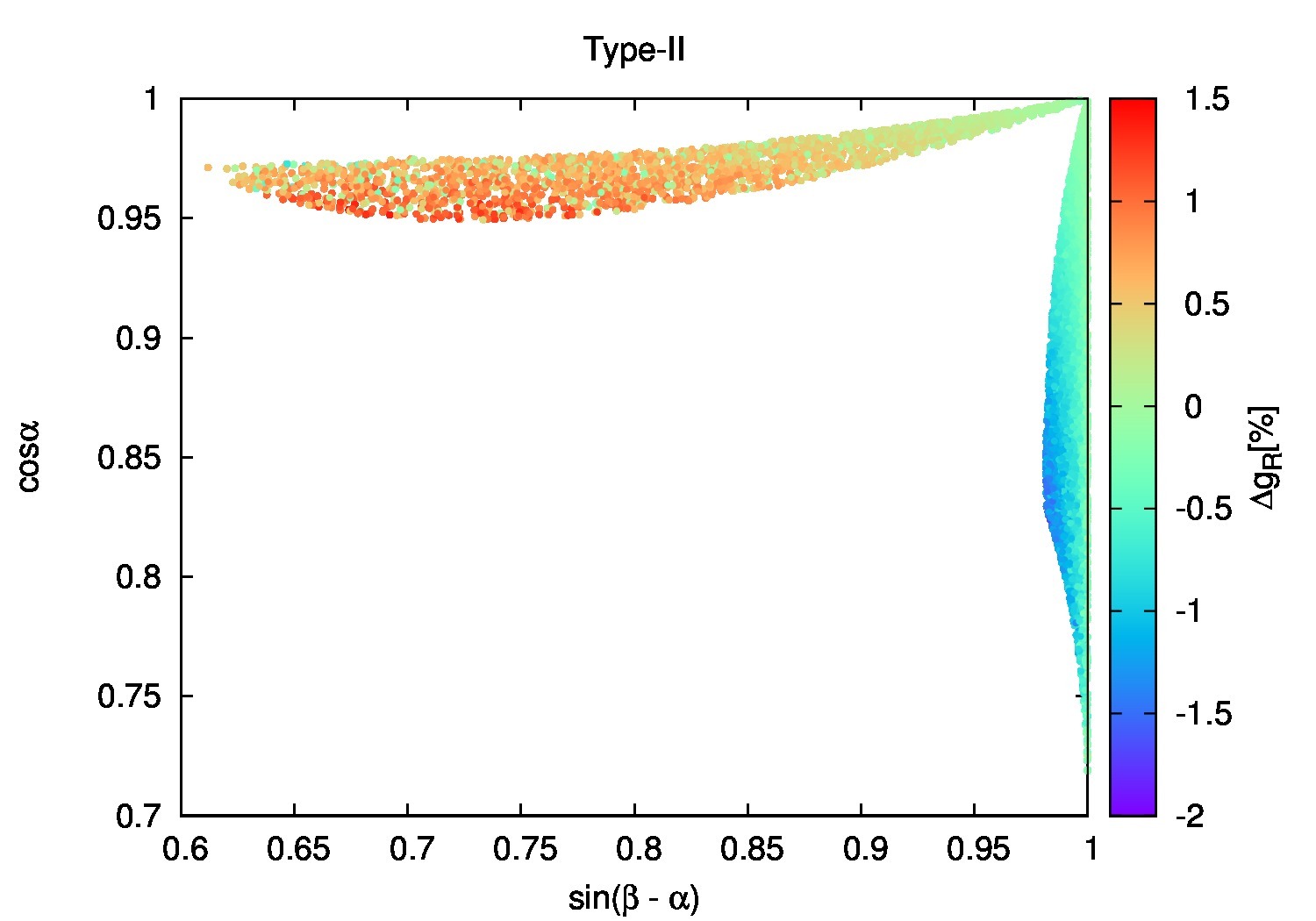}
\caption{\label{gR-THDM} Relative contribution $\Delta g_R$ in 
2HDM type-I  (upper panels)
and type-II (lower panels)
shown as a scatter plot in the ($t_\beta$, $\kappa_d^h$) plan (left), 
($m_{H^0},m_{A^0}$) plan (middle) and ($s_{\beta-\alpha}, c_\alpha$) plan (right)}
\end{figure}
In figure~\ref{gR-THDM}, we plot the correction 
to $g_R$ in 2HDM type-I (upper panels) and 2HDM type-II (lower panels). 
In 2HDM type-I, one can see that the corrections can reach $-5\%$
for $\tan \beta \sim 1$ and the enhancement attains $1\%$. However, in
type-II 2HDM, the suppression of the correction to $g_R$ with respect
to the SM result is smaller and the enhancement is quite bigger than
type-I 2HDM, i.e. $\max{(\Delta g_R)} \simeq 1.5\%$ and $\min{(\Delta
  g_R)} \simeq -2\%$.
\begin{figure}[h]
\centering 
\includegraphics[width=.32\textwidth]{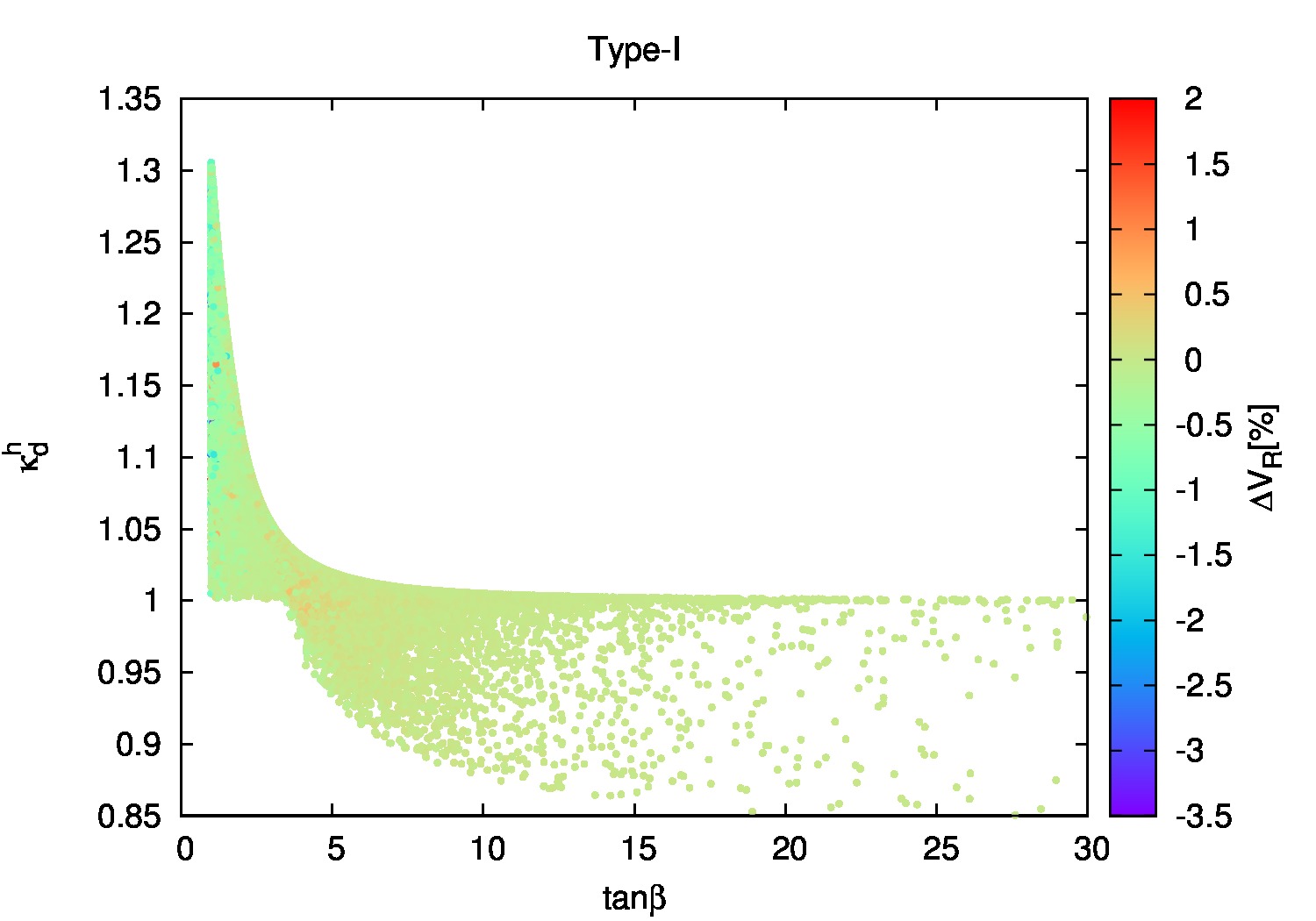}
\hfill
\includegraphics[width=.32\textwidth]{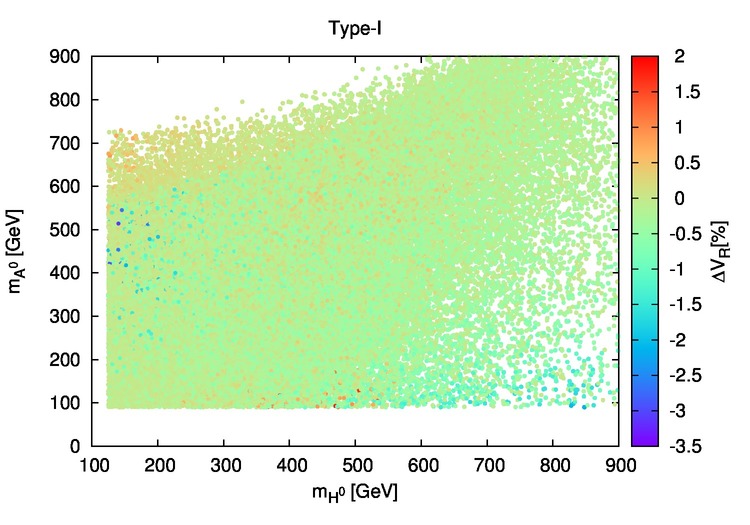}
\hfill
\includegraphics[width=.32\textwidth]{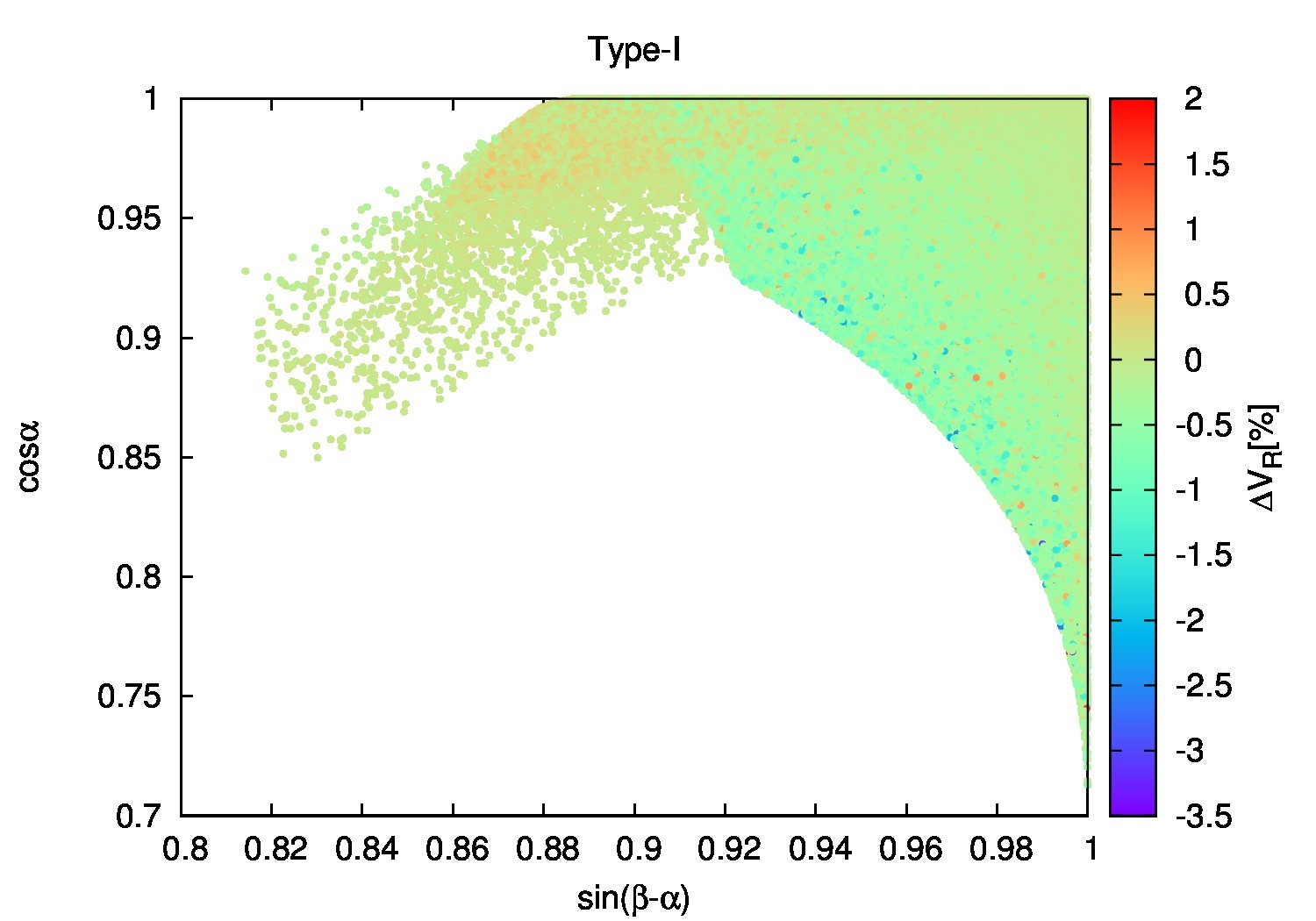}
\hfill
\includegraphics[width=.32\textwidth]{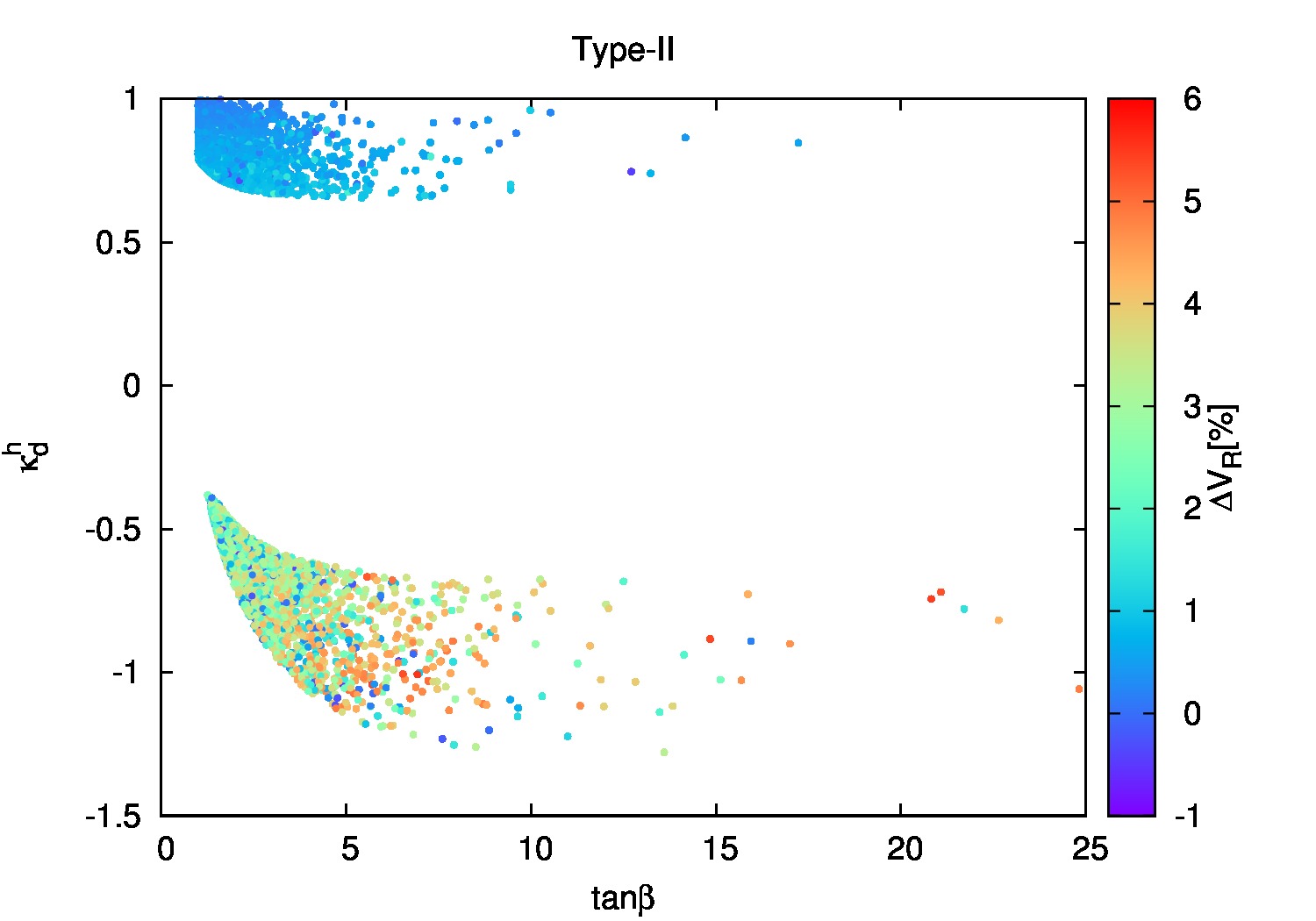}
\hfill
\includegraphics[width=.32\textwidth]{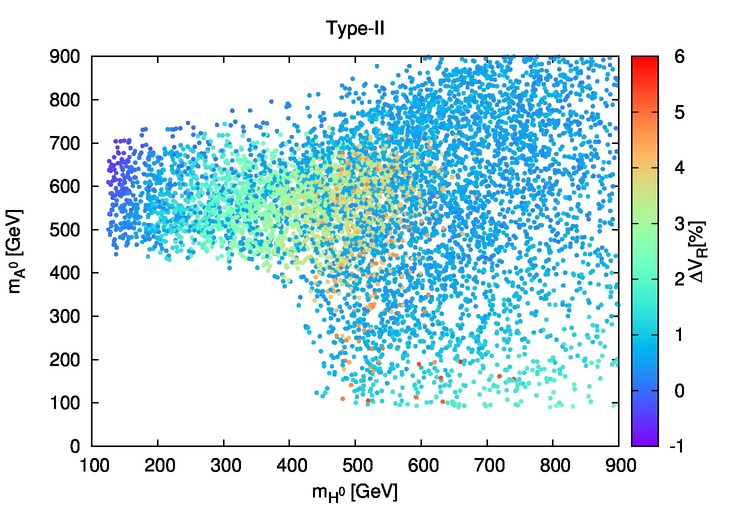}
\hfill
\includegraphics[width=.32\textwidth]{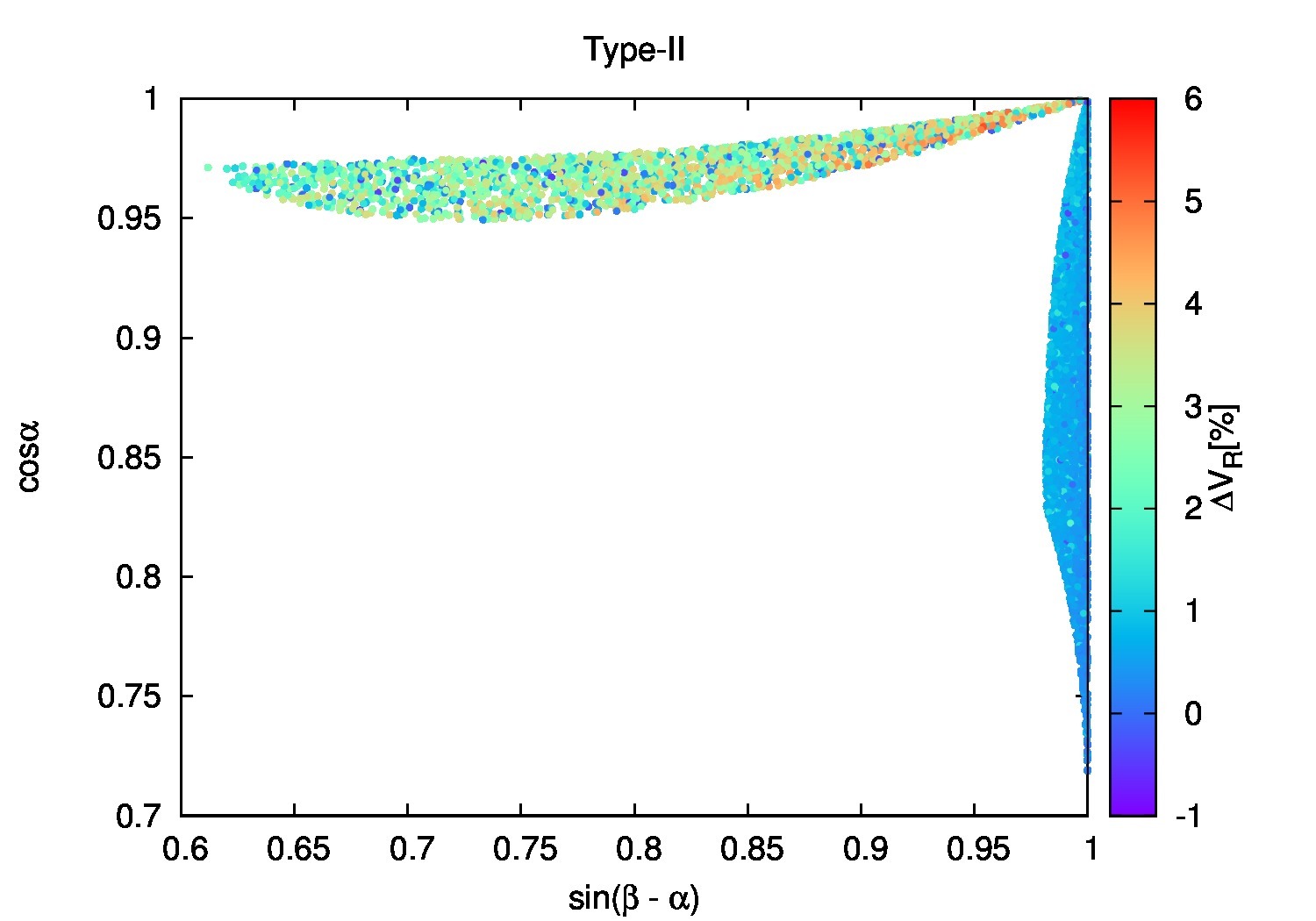}
\caption{\label{VR-THDM} Relative contribution $\Delta V_R$ in 
2HDM type-I  (top) and 2HDM type-II (bottom)
shown as a scatter plot in ($t_\beta$, $\kappa_d^h$) plan (left), 
($m_{H^0},m_{A^0}$) plan (middle) and ($s_{\beta-\alpha}, c_\alpha$) plan (right)}
\end{figure}

In figure \ref{VR-THDM} (upper panels), we illustrate the correction to $V_R$
in 2HDM type-I. It is clear from these plots that the corrections
hardly reach $2\%$  while the maximum of suppression is about $-3.5\%$.
The decoupling limit where $V_R^{2HDM}=V_R^{SM}$ is attained for large
$\tan \beta$, large masses $m_{H^0,A^0}>700 \text{ GeV}$ and 
for $s_{\beta-\alpha}\simeq 1$. 

In the lower panels of figure \ref{VR-THDM}, we can see thatat,
contrarily to 2HDM type-I, the enhancement of $V_R$ in 2HDM type-II is larger 
 and reaches $6\%$ for $m_{H^0} \text{ and } m_{A^0} \in [300:400] \text{
   GeV}$ while the suppression is hardly fulfilled and reaches only $-1\%$. 
\begin{figure}[tbp]
\centering 
\includegraphics[width=.32\textwidth]{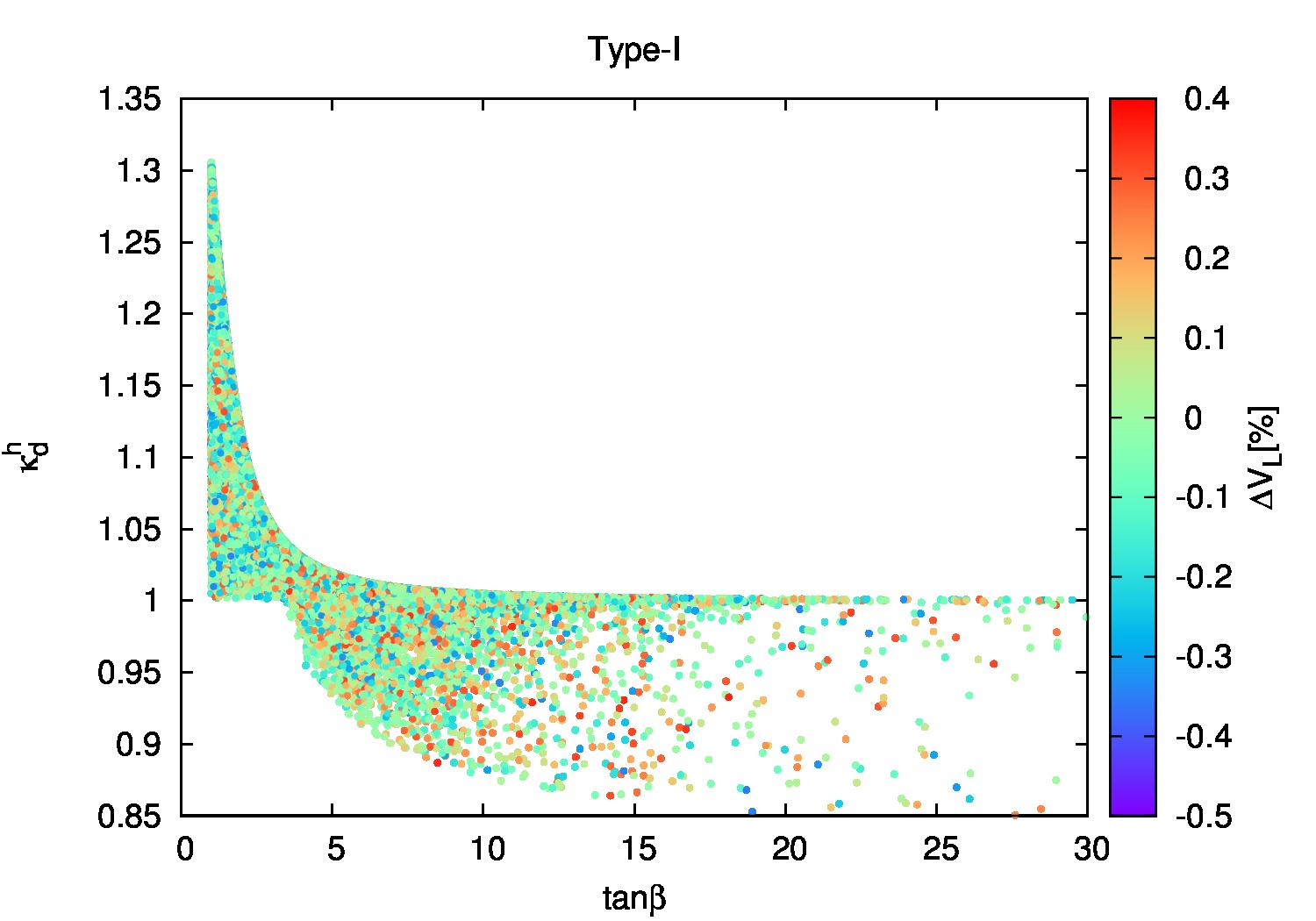}
\hfill
\includegraphics[width=.32\textwidth]{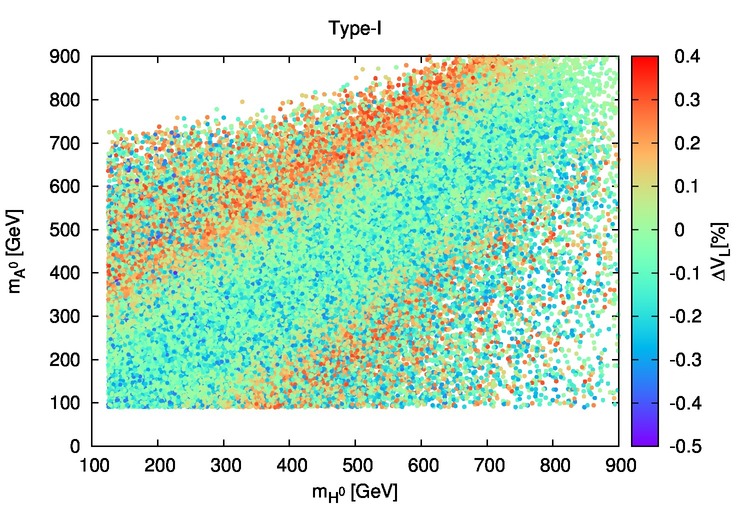}
\hfill
\includegraphics[width=.32\textwidth]{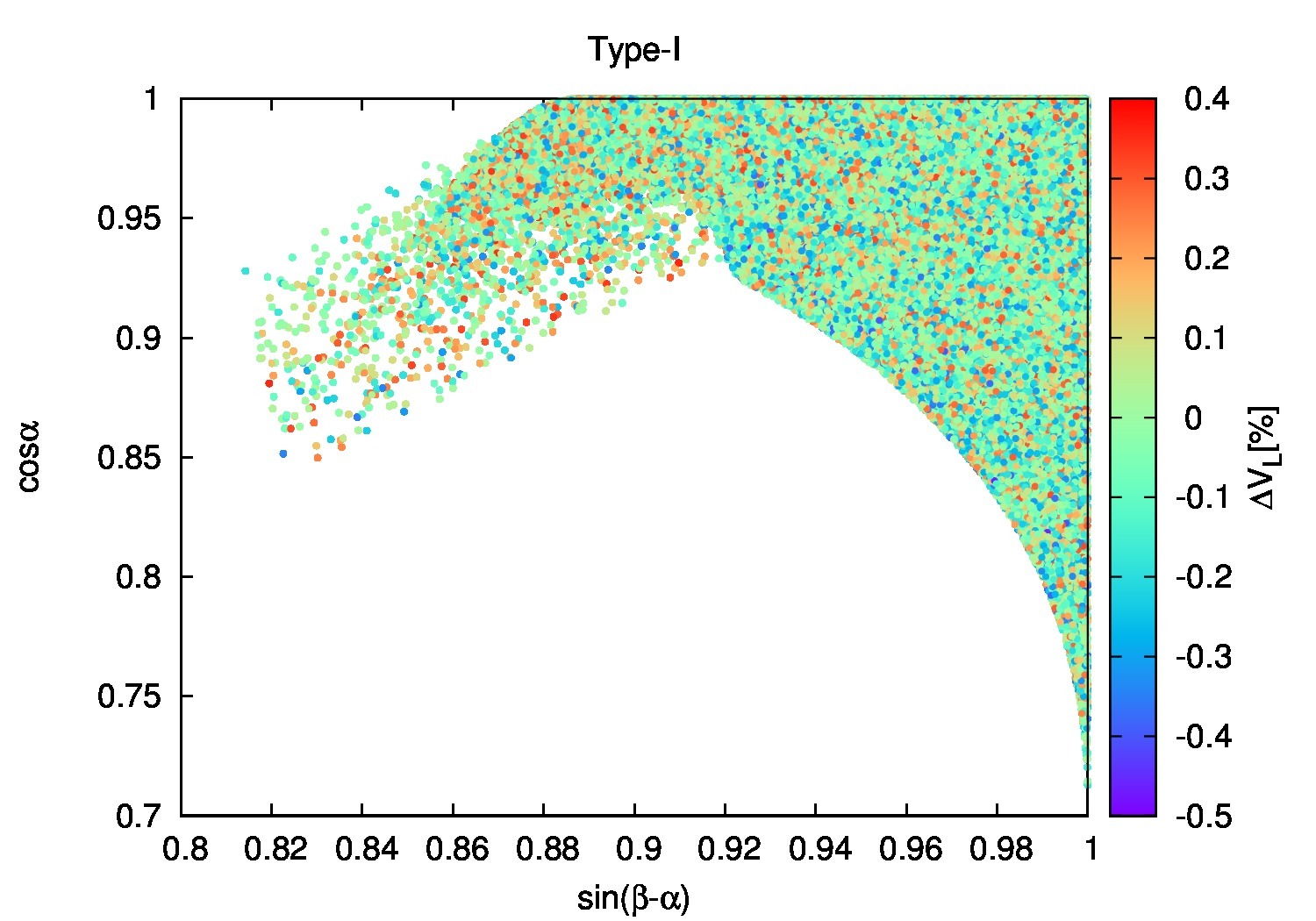}
\hfill
\includegraphics[width=.32\textwidth]{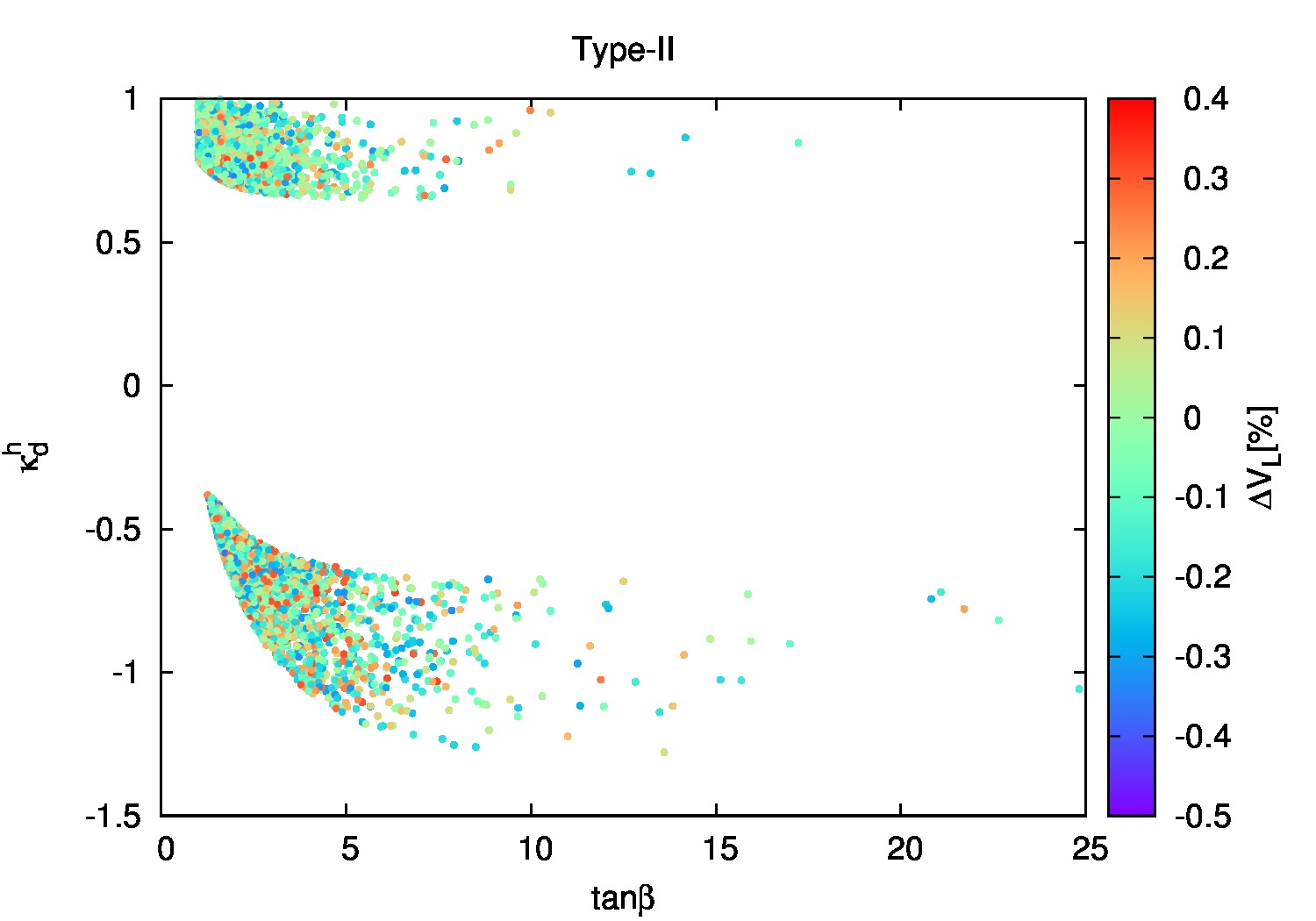}
\hfill
\includegraphics[width=.32\textwidth]{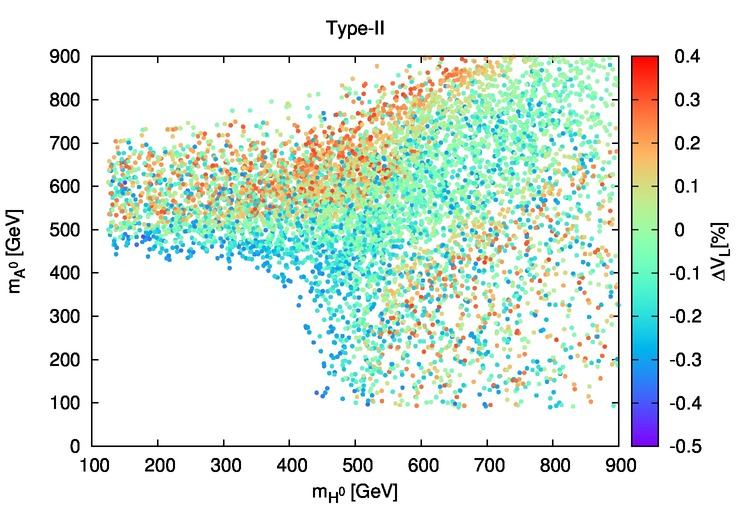}
\hfill
\includegraphics[width=.32\textwidth]{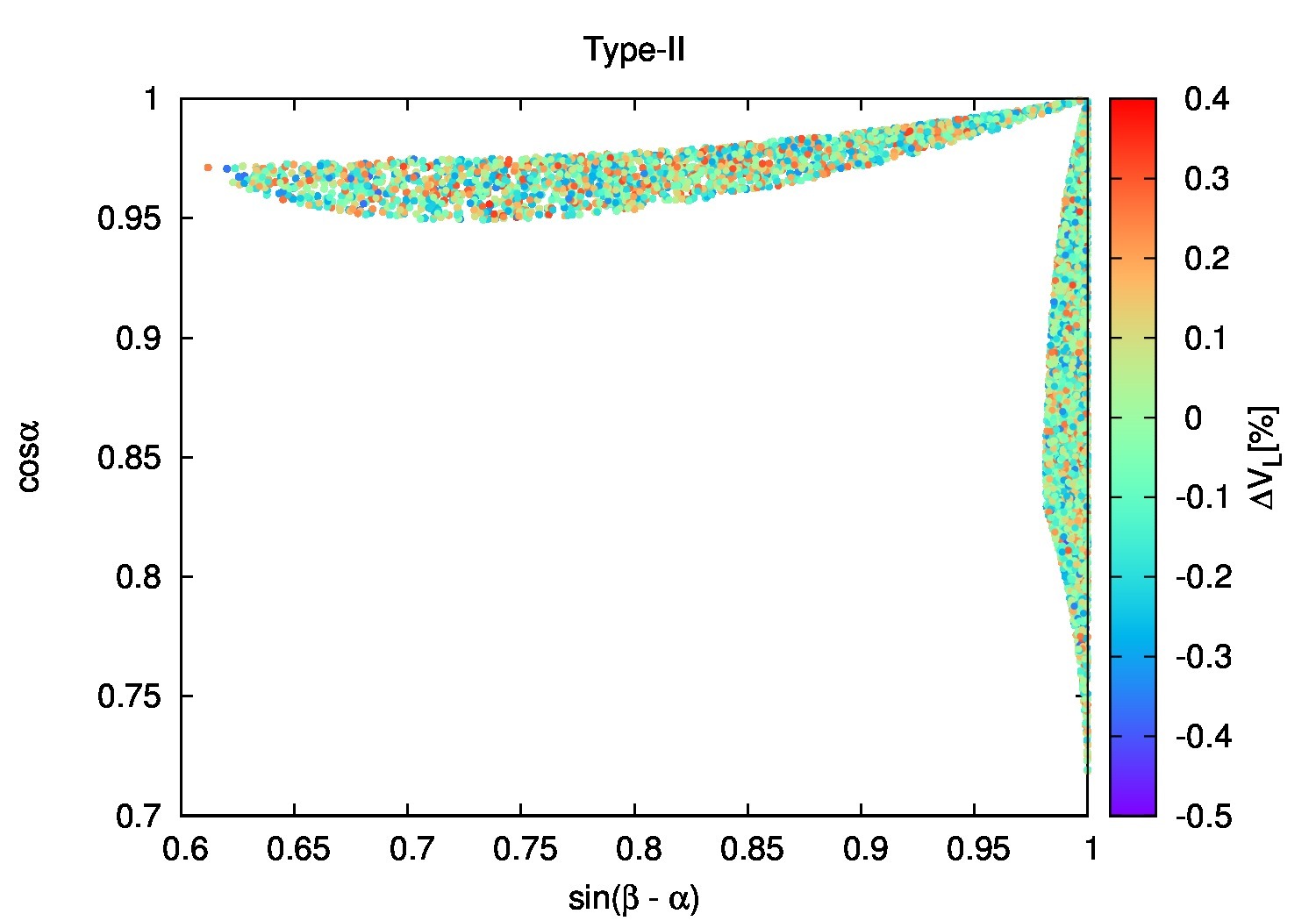}
\caption{\label{VL-THDM} Relative contribution $\Delta V_L$ 
in 2HDM type-I (top) and 2HDM type-II (bottom)
shown as a scatter plot in ($t_\beta$, $\kappa_d^h$) plan (left), 
($m_{H^0},m_{A^0}$) plan (middle) and ($s_{\beta-\alpha}, c_\alpha$) plan (right)}
\end{figure}
In figure \ref{VL-THDM}, we have shown corrections
to the left chiral coupling $V_L$. We see that the corrections are very small
(not exceeding $0.4\%$) in most regions of the parameter space. 
The suppression of $V_L$ with respect to its SM value reaches 
$-0.5\%$ in  the 2HDM type-I and II.
\begin{figure}[tbp]
\centering 
\includegraphics[width=.32\textwidth]{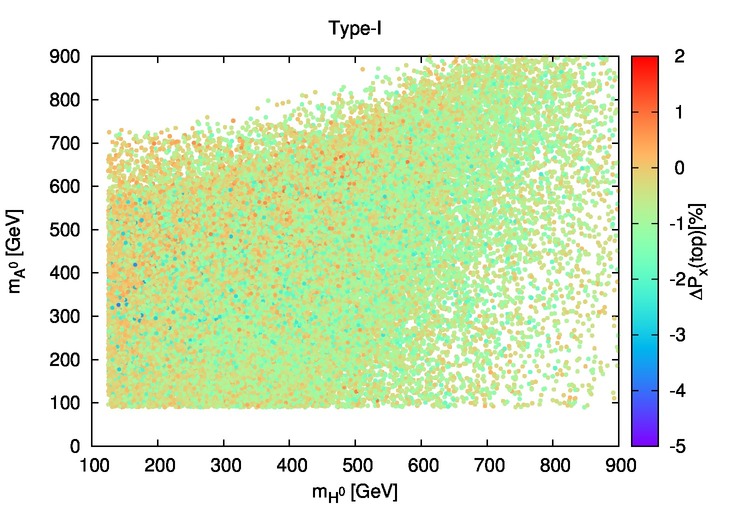}
\hfill
\includegraphics[width=.32\textwidth]{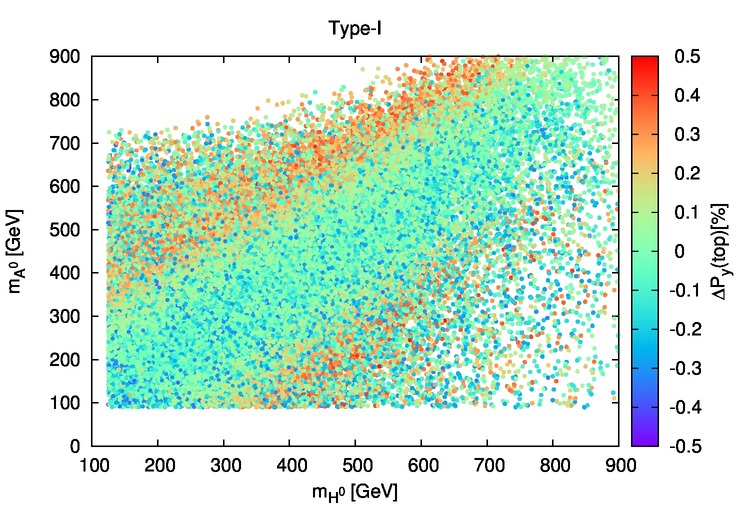}
\hfill
\includegraphics[width=.32\textwidth]{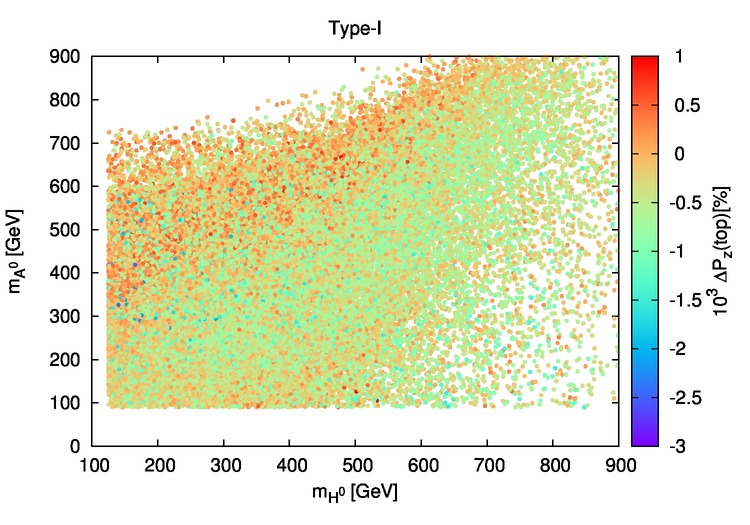}
\hfill
\includegraphics[width=.32\textwidth]{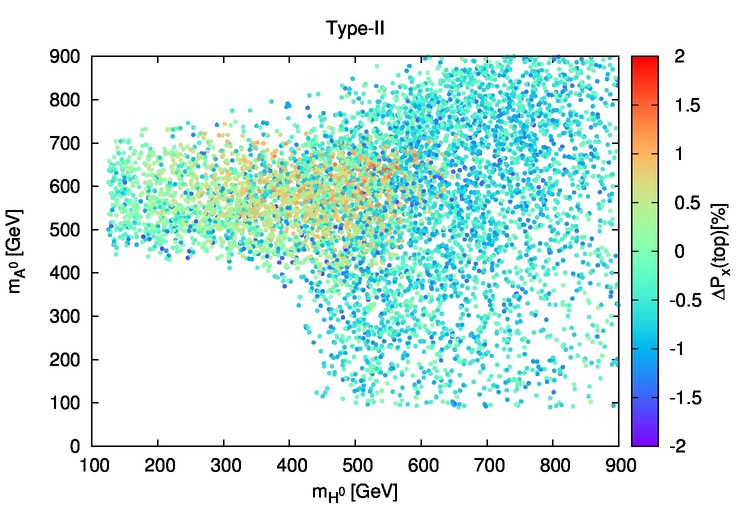}
\hfill
\includegraphics[width=.32\textwidth]{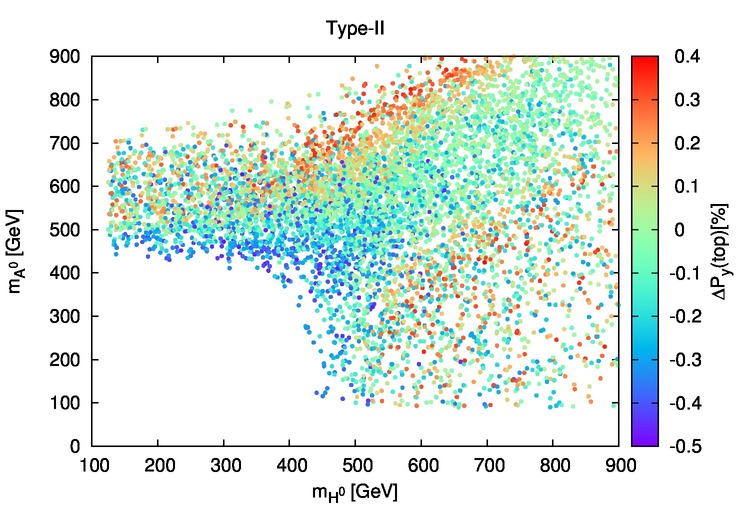}
\hfill
\includegraphics[width=.32\textwidth]{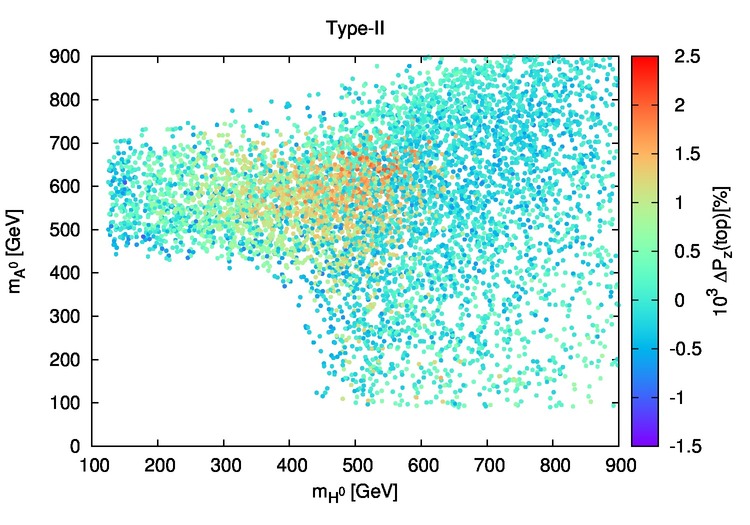}
\caption{\label{Polarisation-THDM} Scatter plots in 
($m_{H^0}$, $m_{A^0}$) plan where 
the palette shows the values of $\Delta P_x$ (left panels), 
$\Delta P_y$ (middle panels) 
and $\Delta P_z$ (right panels) in the 2HDM type-I (upper panels) and 
 2HDM type-II (lower panels)}
\end{figure}

\subsection{Top Polarization}
We also studied numerically the top polarization in the 
channel $qg \to q' t \bar{b}$ at the LHC for 
 $\sqrt{s}=14 \text{ TeV}$.
The relative correction is defined as:
\begin{eqnarray}
 \Delta P_i = \frac{P_i^{\text{2HDM}}-P_i^{\text{SM}}}{P_i^{\text{SM}}} \quad, \quad i=x,y,z
 \label{polarisation}
\end{eqnarray}
Following Ref.~\cite{Aguilar-Saavedra:2014eqa},
the axes were defined as follow; $z$ axis is the direction of the spectator
quark $q'$, the $y$-axis is orthogonal to the direction of the momentum 
of the initial quark $q$ and the momentum of the spectator quark 
$q'$ and the $x$-axis is chosen such that the system is right handed. 
We have taken the expressions of the components of the polarization 
vector from \cite{Aguilar-Saavedra:2014eqa} for both the top and
anti-top quarks at $14 \text{ TeV}$. 
We quote the results for the case of the SM in table.~(\ref{Pol-SM}).
\begin{table}[!h]
\begin{center}
\begin{tabular}{|l||l||l|}
 \hline \hline
 & Top quark & Anti-top quark \\ \hline
 $P_x$ & $-0.0179074$ & $-0.107771$ \\ \hline
 $P_y$ & $0.0040848$ & $9.66629 \times 10^{-6}$ \\ \hline
 $P_z$ & $0.880908$ & $-0.850601$ \\ \hline \hline
\end{tabular}
\end{center}
\caption{Values of the polarization of $t/\bar{t}$ in the SM 
 at $\sqrt{s}=14$ TeV.  Formula are taken from 
Ref.~\cite{Aguilar-Saavedra:2014eqa}}
\label{Pol-SM}
\end{table}
In figure \ref{Polarisation-THDM} (upper panels), we plot the 
relative correction  $\Delta P_i, i=x,y,z$
in the 2HDM type-I for the top quark in the ($m_{H^0}$, $m_{A^0}$) plan.
We see that $\Delta P_x$ reaches $2\%$ as a maximum of enhancement. 
The suppression of $P_x$ with respect to the SM value reaches $-5\%$. 
Corrections to $P_y$ are shown 
in the middle panel of figure \ref{Polarisation-THDM}, 
the corrections are rather smaller than 
those corresponding to $P_x$: $\max \Delta P_y \sim 0.5\%$ and 
the suppression is of order of $-0.5\%$ which implies that non 
significant deviation from the 
SM is attained. We note also that the corrections to $P_z$ 
(right panel of figure \ref{Polarisation-THDM})
are even more smaller ($0.001\%$ as a maximum). \\
In figure~\ref{Polarisation-THDM} (lower panels), we plot the 
relative corrections to the components of the polarization vector
of the top in 2HDM type-II. We see that in this model, 
corrections are very small.
$\max\{\Delta P_x,\Delta P_y,\Delta P_z\} = \{2\%,0.4\%,0.0025\%\}$
and $\min\{\Delta P_x,\Delta P_y,\Delta P_z\} = \{-2\%,-0.5\%,0.0015\%\}$. 

\subsection{W helicity fractions}
Anomalous $tbW$ couplings could be probed by measuring the 
$W$ boson helicity fractions in the top quark decay (unpolarized decay)
\cite{Chatrchyan:2013jna,Aad:2012ky}.
These polarization states are proven to be sensitive to 
new physics effects \cite{Kane:1991bg} where the $W$ boson could 
be produced with positive ($R$), negative ($L$) 
or zero ($0$) helicity states, $\Gamma(t\to b W^+) = 
\Gamma_L + \Gamma_R + \Gamma_0$. 
Expressions of the polarized widths in terms of 
the anomalous couplings are taken from \cite{AguilarSaavedra:2006fy}. 
The polarization of the $W$ boson could be measured by looking 
to the angular distributions 
of its decay products (especially into leptons). 
The differential decay rate of the unpolarized top quark is given by:
\begin{eqnarray}
 \frac{1}{\Gamma}\frac{d\Gamma}{d\cos\theta_l^*} = \frac{3}{8} (1+\cos\theta_l^*)^2 F_R + 
 \frac{3}{8} (1-\cos\theta_l^*)^2 F_L + \frac{3}{4} \sin^2\theta_l^* F_0,
 \label{helicity-W}
\end{eqnarray}
with$F_i=\Gamma_i/\Gamma$ are the helicity fractions 
and $\theta_l^*$ is the angle between
the lepton three-momentum in the rest frame of the parent $W$ boson and 
the $W$ boson three momentum in the top quark rest frame. The SM predictions 
are known up to NNLO in QCD \cite{Czarnecki:2010gb}; 
$$F_0=0.687\pm0.005, F_L=0.311\pm0.005\qquad  \text{and} \qquad
F_R=0.0017\pm0.0001$$ Calculations of the helicity fractions have been performed 
in the framework of the MSSM in \cite{Cao:2003yk}. \\

For numerical analysis, we define the ratios $\delta F_i$ by :
\begin{eqnarray}
 \delta F_i = \frac{F_i^{\text{2HDM}} - F_i^{\text{SM}}}{F_i^{\text{SM}}}
 \label{helicity-ratio}
\end{eqnarray}
where $F_i^\text{SM}$ includes complete one-loop corrections 
(in $\alpha_s$ and $\alpha$)
while $F_i^\text{2HDM}$ contains additional the contribution from the extra
particles of the 2HDM and its interference with the pure SM (EW and QCD) contribution. \\
\begin{figure}[tbp]
\centering 
\includegraphics[width=.32\textwidth]{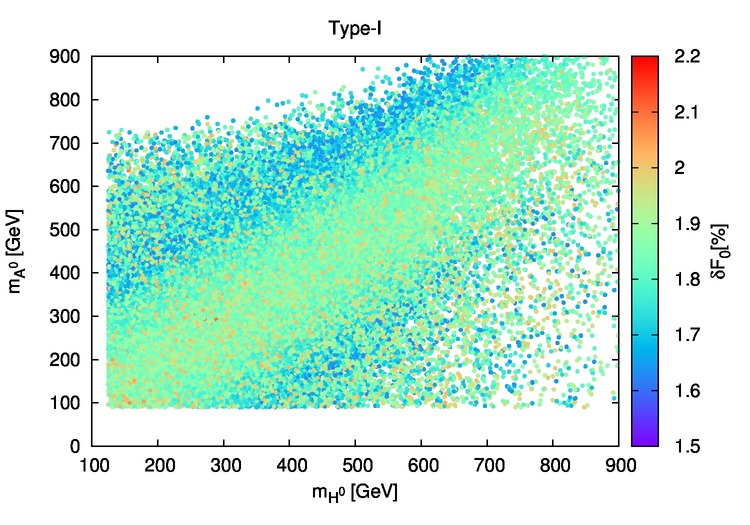}
\hfill
\includegraphics[width=.32\textwidth]{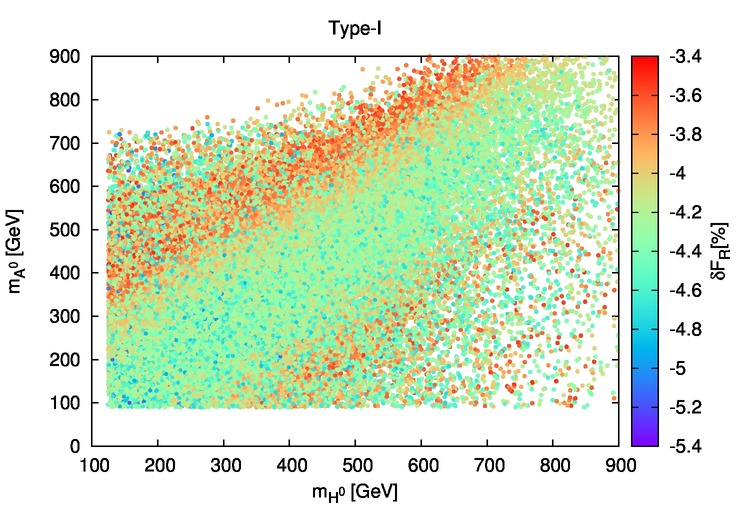}
\hfill
\includegraphics[width=.32\textwidth]{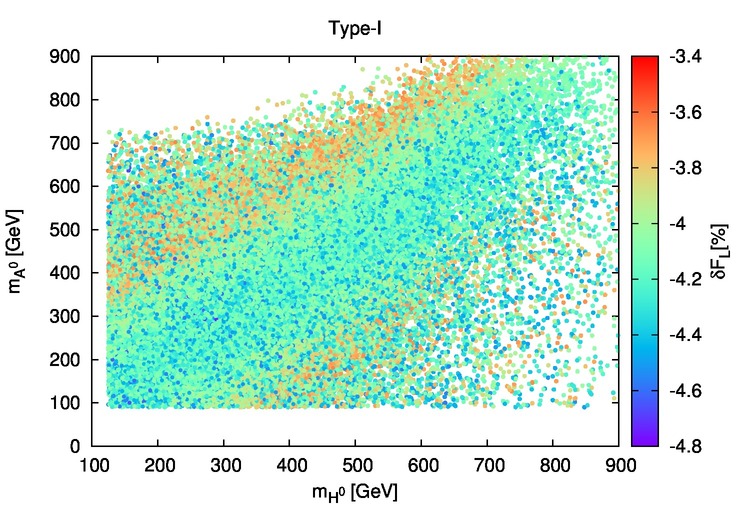}
\hfill
\includegraphics[width=.32\textwidth]{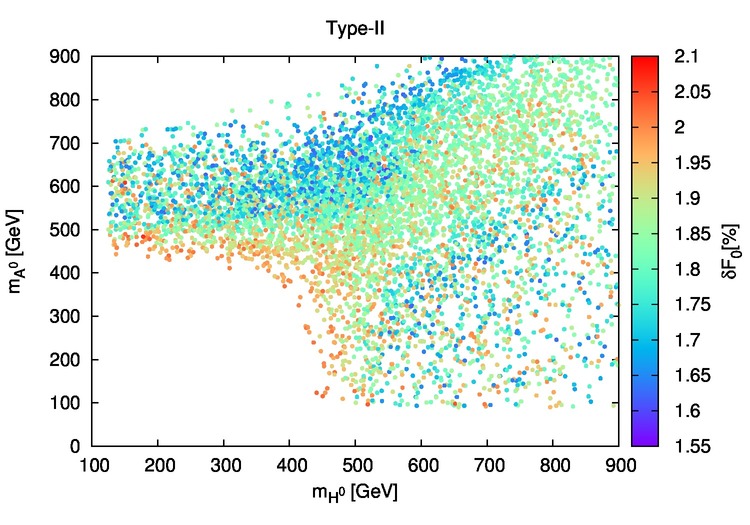}
\hfill
\includegraphics[width=.32\textwidth]{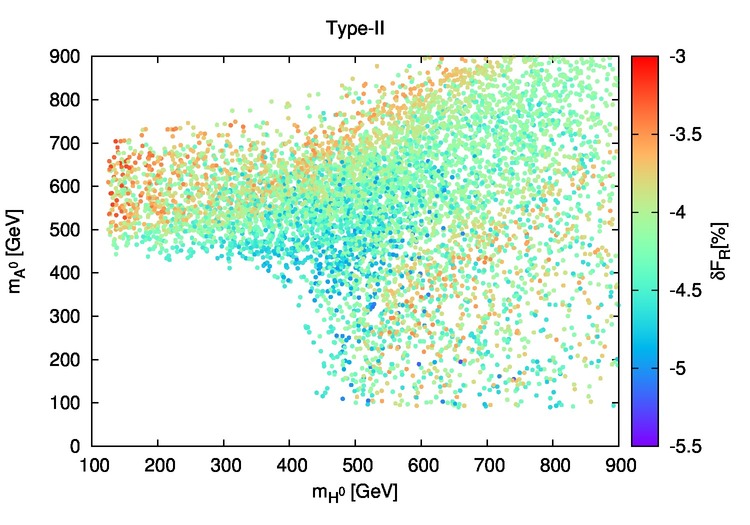}
\hfill
\includegraphics[width=.32\textwidth]{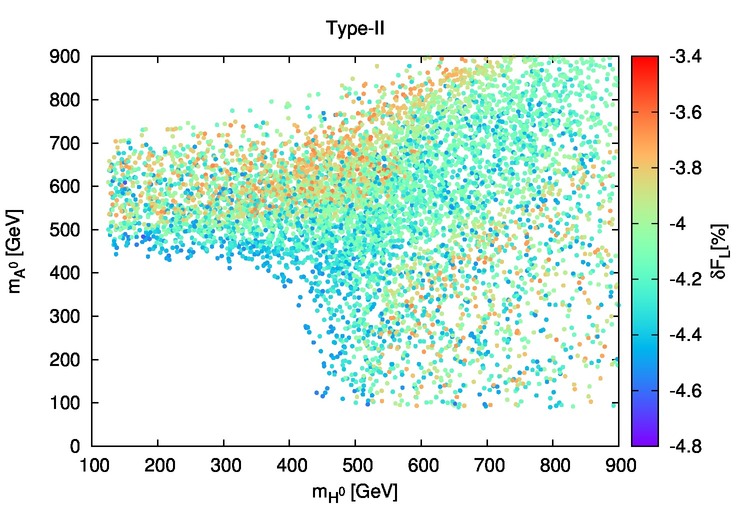}
\caption{\label{Helicity-THDM} Scatter plot in the ($m_{H^0}$, $m_{A^0}$) 
plan where the palettes show the values of $\delta F_0$ (left panel), 
$\Delta F_R$ (middle panel) and $\Delta F_L$ (right panel) 
in the 2HDM type-I (top) and 2HDM type-II (bottom)}
\end{figure}
In figure.~\ref{Helicity-THDM} (upper panels), we plot the contribution to 
the helicity fractions in 2HDM type-I. $\delta F_0$ is shown in the left-panel,
where we observe that the corrections are quite small $\max \delta F_0 = 2.2\%$
and always enhancing $F_0$ with respect to its SM value while the minimum 
of the correction is $1.5\%$. Corrections to $F_R$ is depicted in 
figure.~\ref{Helicity-THDM} (middle panel), 
we see that the corrections are always suppressing the 
SM value; $\max \delta F_R\sim -3.4\%$ and $\min \delta F_R\sim -5.4\%$. 
We notice that $F_R$ is very small and vanishes in the limit 
$m_b \to 0$. 
We illustrate in the right panel of the same figure the 
correction to $F_L$. As it can be seen, there is 
always suppression of $F_L$ with respect to its SM value 
$-4.8\% \leq \delta F_L \leq -3.4\%$.\\
In figure \ref{Helicity-THDM} (lower panels), 
corrections to $F_i$ are shown in 
2HDM type-II. In the left panel, we see that correction to $F_0$ is more or less
of the same size as for the case of 2HDM type-I. 
The maximum of $\delta F_0$ is $2.1\%$
reached where the masses are quite small 
$m_{H^0, A^0} \sim 300-400 \text{ GeV}$. In the middle 
panel of figure (\ref{Helicity-THDM}), we show the correction to $F_R$. 
We see that $F_R$ 
is always suppressed with respect to its SM value. 
$-5.5\% \leq \delta F_R \leq -3\%$. Finally,
the extra contribution to $F_L$ is shown in the right 
panel of figure. \ref{Helicity-THDM}, 
one can see that the corrections are the same in 2HDM type-I and type-II,
 e.g $-4.8\% \leq \delta F_R \leq -3.4\%$ while the maximum of suppression
is reached for the region of low scalar masses.
\section{Conclusion} 
\label{Conclusion}

We have computed the complete one loop 
contribution to the anomalous $tbW$ couplings 
in the 2HDM. We give for the first time the analytical expressions of the
anomalous couplings in terms of the Passarino Veltman functions. 
We have evaluated both the anomalous couplings $g_L$
and $g_R$ as well as left handed $V_L$ and right handed $V_R$ 
component of $tbW$. The computation is done by diagrammatic method 
in the Feynman gauge using dimensional regularization in the 
On-shell renormalization scheme. 

We show sensitivity of the 2HDM parameters to the various anomalous $tbW$
couplings taking into account recent LHC constraints. 
We also illustrate the overall sensitivity to the 2HDM parameters 
to some LHC observables such as: top polarization in single top production
through t-channel as well as $W^\pm$ helicity fractions in top decay.
We also project our numerical results on $\kappa_D$, which is the Yukawa
coupling of the Higgs to down quarks and also on $\sin(\beta-\alpha)$ which
leasure departure from decoupling limit of 2HDM.

The effect on most of the observales we consider are rather small.
It will be rather a difficult task to disentangle the 2HDM from SM even with
the High luminosity LHC option. However, with the projected 
 Super B Factory experiments with high luminosity, 
from  the precise measurement of $b\to s\gamma$ we would have a 
strong limit  on $V_R$ and $g_L$.

\acknowledgments  This work was supported by the Moroccan Ministry of Higher
Education and Scientific Research MESRSFC and  CNRST: 
"Projet dans les domaines prioritaires
de la recherche scientifique et du d\'eveloppement technologique": PPR/2015/6.
A.J would like to thank the STEP Programme (ICTP-Trieste) and 
GDRI P2IM Maroc-France (LAPTh, CNRS)
for financial support 
during his stay where part of this work has been done.
The authors would like to thank Fawzi Boudjema for careful 
reading of the manuscript.
\appendix
\section{Anomalous Tensor couplings in the SM}
\label{appen-1}
In this appendix, we give our numerical results for the anomalous
tensor couplings in the SM and compare with results from 
\cite{GonzalezSprinberg:2011kx} and
\cite{Gonzalez-Sprinberg:2015dea}.

In table (\ref{Compa.1}) we show the values of $g_L$. 
In most of the cases, there is an agreement between our results and 
those presented by the authors 
of \cite{GonzalezSprinberg:2011kx}
except for diagrams with $b W^+ Z$ and $b G^+ Z$ exchange where our 
results are two times larger. 
On the other hand,
in the diagrams with $b W^+ H$ and $b G^+ G^0$ exchange,
we have found that our imaginary part of $g_L$ has a 
different sign to that found 
in \cite{GonzalezSprinberg:2011kx}. \\
\begin{table}[!h]
\begin{center}
 \begin{tabular}{||l|l|l||}
 \hline 
 Diagram & Contribution to $g_L$ & Vidal et al. \cite{GonzalezSprinberg:2011kx} \\ \hline
 $t Z W^-$ & $-0.0147$ & $-0.0141$ \\ \hline
 $t H W^-$ & $0$  & $0$  \\ \hline
 $t G^0 G^-$ &  $-0.00532$  & $-0.0051$   \\ \hline
 $t G^- H$ & $-0.010$ & $-0.0088$ \\ \hline
 $t Z G^-$ & $-0.0016$ & $-0.0012$  \\ \hline
 $t \gamma W + t \gamma G^-$ & $-0.00925$ & $-0.0094$  \\ \hline
 $b W^+ Z$ & $-0.042 - 0.0457 i$ & $-0.0201-0.0214 i$ \\ \hline
 $b W^+ H$ & $0.0089 + 0.0155 i$ & $0.0086-0.0120 i$  \\ \hline
 $b G^+ G^0$ & $-0.0033 + 0.0172 i$  & $-0.0029 - 0.0167i$  \\ \hline
 $b G^+ H$ & $-0.000356 - 0.0138 i$ & $-0.0019 + 0.0111 i$  \\ \hline
 $b G^+ Z$ & $-0.000765 - 0.000555 i$ & $-0.00039 - 0.00028 i$  \\ \hline
 $b W^+ \gamma + b G^+ \gamma$ & $-0.0262 + 0.0241 i$ & $-0.0270 + 0.0250 i$  \\ \hline
 $Z t b$ & $-0.006846$ &  $-0.0067$ \\ \hline
 $\gamma t b$ & $0.011181$ & $0.0115$  \\ \hline
 $G^0 t b$ & $-0.01134$ & $-0.0109$ \\ \hline
 $H t b$ & $-0.0162$ & $-0.0153$ \\ \hline
 $\Sigma(EW)$ & $-0.128529 - 0.00330156i$ & $-0.102 - 0.0014 i$ \\ \hline 
 $g t b$ & $-1.10326$ & $-1.12$  \\
 \hline 
 \end{tabular}
\caption{A comparison between our results and those of \cite{GonzalezSprinberg:2011kx} corresponding to $10^3 g_L$}
\label{Compa.1}
 \end{center}
\end{table}
The table (\ref{Compa.2}) shows the values of coupling $g_R$ for different
diagrams and compares with the results of Vidal \cite{GonzalezSprinberg:2011kx}.
One can see that the same remarks apply here as for the case of tensor coupling $g_L$. 
\begin{table}[tbp]
\begin{center}
 \begin{tabular}{||l|l|l||}
 \hline 
 Diagram & Contribution to $g_R$ & J.Vidal et al\\ \hline
 $t Z W^-$ & $-1.211$ & $-1.176$ \\ \hline
 $t H W^-$ & $0.26147$  & $0.220$  \\ \hline
 $t G^0 G^-$ &  $-0.3644$  & $-0.344$  \\ \hline
 $t G^- H$ & $0.56$ & $0.462$  \\ \hline
 $t Z G^-$ & $-0.02949$ & $-0.050$  \\ \hline
 $t \gamma W + t \gamma G^-$ & $0.5706$ & $0.572$  \\ \hline
 $b W^+ Z$ & $-1.33481 - 1.46899 i$ & $-0.623 -0.664 i$  \\ \hline
 $b W^+ H$ & $0$ & $0$ \\ \hline
 $b G^+ G^0$ & $0.0001675 - 0.0011 i$  & $(1.5 + 11 i) \times 10^{-4}$  \\ \hline
 $b G^+ H$ & $ -0.000439 - 0.00117 i $ & $(-4.3 + 8.6 i) \times 10^{-4}$  \\ \hline
 $b G^+ Z$ & $-0.1820 - 0.132 i $ & $-0.088-0.062 i$  \\ \hline
 $b W^+ \gamma + b G^+ \gamma$ & $ 0.118 - 0.503 i$ & $0.0114 - 0.509 i$  \\ \hline
 $Z t b$ & $-0.4096$ &  $-0.397$ \\ \hline
 $\gamma t b$ & $0.0669$ & $0.068$ \\ \hline
 $G^0 t b$ & $-0.00069$ & $-6.8 \times 10^{-4}$ \\ \hline
 $H t b$ & $-0.00077$ & $-6.2 \times 10^{-4}$ \\ \hline
 $\Sigma(EW)$ & $-1.95628 - 2.10655 i$ & $-1.24 - 1.23 i$  \\ \hline 
 $g t b$ & $-6.60729$ & $-6.61$  \\
 \hline 
 \end{tabular}
\caption{A comparison between our results and those of \cite{GonzalezSprinberg:2011kx} correponding to $10^3 g_R$}
\label{Compa.2}
 \end{center}
\end{table}

\begin{table}[!h]
 \begin{center}
 \begin{tabular}{||l|l|l||}
  \hline
  Diagram & Contribution to $V_R$ & Result of \cite{Gonzalez-Sprinberg:2015dea} \\ \hline
  $t Z W^\pm$ & $ 2.18162\times 10^{-5}$  & $2.01 \times 10^{-5}$\\ \hline
  $t \gamma W^\pm$ & $-1.22114 \times 10^{-5}$ & $-1.10 \times 10^{-5}$\\ \hline
  $t H W^\pm$ & $0$ & $0$ \\ \hline
  $t G^\pm G^0 + t H G^\pm$ & $-1.67866\times 10^{-5}$  & $-1.55 \times 10^{-5}$ \\ \hline
  $t Z G^\pm$ & $0.117165 \times 10^{-5}$ & $0.1 \times 10^{-5}$\\ \hline
  $t \gamma G^\pm$ & $0.76815\times 10^{-5}$ & $0.69 \times 10^{-5}$\\ \hline
  $b W^\pm Z$ & $(1.19335 + 8.90489 i) \times 10^{-4}$ & $(1.12 + 8.24 i)\times 10^{-5}$\\ \hline
  $b W^\pm \gamma$ & $(8.97983 - 4.71769 i) \times 10^{-5}$ & $(8.34 - 4.25 i) \times 10^{-5}$  \\ \hline
  $b W^\pm H$ & $0$ & $0$\\ \hline
  $b G^\pm G^0 + b G^\pm H$ & $(1.05897 + 1.9014 i) \times 10^{-5}$  & $(1.01 - 0.35i)\times 10^{-5}$ \\ \hline
  $b G^\pm Z$ & $(0.00109755 +  0.360717 i) \times 10^{-5}$  & $0.31 i \times 10^{-5}$\\ \hline
  $b G^\pm \gamma$ & $(-4.82503 + 2.5363 i) \times 10^{-5}$  & $(-4.47 + 2.29 i)\times 10^{-5}$\\ \hline
  $Z t b$ & $-2.5271 \times 10^{-5}$ & $-2.30 \times 10^{-5}$\\ \hline 
  $\gamma t b$ & $-2.98898\times 10^{-5}$  & $-2.78 \times 10^{-5}$ \\ \hline
  $G^0 t b + H t b$ & $-1.13206 \times 10^{-5}$  & $-1.03 \times 10^{-5}$\\ \hline
  $\Sigma(\text{EW})$ & $(-0.0727959 + 8.98568 i) \times 10^{-5}$ & $(0.06 + 6.23i) \times 10^{-5}$\\ \hline
  $g t b$ & $2.91224 \times 10^{-3}$ & $2.68 \times 10^{-3}$\\ \hline
 \end{tabular}
 \caption{The right chiral coupling $V_R$ in the SM at the one-loop order}
 \label{VR-SM}
\end{center}
\end{table}
In table (\ref{VR-SM}), we show contribution to $V_R$ for 
different diagrams in the SM and
compare with the recent results of Vidal et al reported in \cite{Gonzalez-Sprinberg:2015dea}. 
We have checked the correctness of our results, for 
certain diagrams where the results are 
not consistent with \cite{GonzalezSprinberg:2011kx,Gonzalez-Sprinberg:2015dea}, 
both by the Feynman parameterization and Passarino-Veltman reduction
methods.
\section{Top Quark Anomalous Couplings $g_L, g_R \textrm{ and } V_R$ in the Two-Higgs-Doublet-Model}
\label{appen-2}
In this appendix, we present for the first time the analytical expressions 
of the anomalous couplings 
for different diagrams in the 2HDM in terms of Passariono-Veltman functions. 
Where $\kappa_d^h$, $\kappa_d^H$  and $\kappa_d^A$ are the Yukawa couplings 
defined in equation.~(\ref{Yukawa-1}) and in table (\ref{Yukawa-2})
\subsection{The Tensorial Coupling $g_L$}
\bea
&&\hspace*{-3cm} g_L^{b h^0 t} = \frac{\alpha c_\alpha m_b m_t^2 \kappa_d^h}{16 M_W \pi s_\beta s_W^2} \{ C_{12} - C_2 \} \\
&&\hspace*{-3cm} g_L^{b H^0 t} = \frac{\alpha s_\alpha m_b m_t^2 \kappa_d^H}{16 M_W \pi s_\beta s_W^2} \{ C_{12} - C_2 \} \\
&&\hspace*{-3cm} g_L^{A^0bt} = \frac{-\alpha m_b m_t^2 \kappa_d^A}{16 M_W \pi t_\beta s_W^2} \{ C_{12} + C_{22} \} \\
&&\hspace*{-3cm} g_L^{b t G^0} = \frac{-\alpha m_b m_t^2}{16 M_W \pi s_W^2} \{C_1 + C_{12} \} \\
&&\hspace*{-3cm} g_L^{b h^0 H^\pm} = \frac{\alpha c_{\beta-\alpha} m_b \kappa_d^h}{16 M_W \pi s_W^2 t_\beta} \{m_b^2 t_\beta \kappa_d^A 
 (C_0 - C_{11} - C_{12} + C_2) - m_t^2 (C_{12} + C_2 + C_{22})\} \\
&&\hspace*{-3cm} g_L^{b H^0 t} = \frac{\alpha s_\alpha m_b m_t^2 \kappa_d^H}{16 M_W \pi s_\beta s_W^2} \{ C_{12} - C_2 \}\\
&&\hspace*{-3cm} g_L^{A^0bt} = \frac{-\alpha m_b m_t^2 \kappa_d^A}{16 M_W \pi t_\beta s_W^2} \{ C_{12} + C_{22} \}\\
&&\hspace*{-3cm} g_L^{b t G^0} = \frac{-\alpha m_b m_t^2}{16 M_W \pi s_W^2} \{C_1 + C_{12} \} \\
&&\hspace*{-3cm} g_L^{h^0H^\pm t} = \frac{\alpha c_\alpha c_{\beta-\alpha} m_b m_t^2}{16 \pi M_W s_\beta t_\beta s_W^2} 
 \{ (-1+t_\beta \kappa_d^A) C_{12} + t_\beta \kappa_d^A (2 C_2 + C_{22}) \}  \\
&&\hspace*{-3cm} g_L^{b h^0 H^\pm} = \frac{\alpha c_{\beta-\alpha} m_b \kappa_d^h}{16 M_W \pi s_W^2 t_\beta} \{m_b^2 t_\beta \kappa_d^A 
 (C_0 - C_{11} - C_{12} + C_2) - m_t^2 (C_{12} + C_2 + C_{22})\} \\
  &&\hspace*{-3cm} g_L^{H^\pm H^0 t} =  \frac{-\alpha s_\alpha s_{\beta-\alpha} m_b m_t^2}{16 \pi M_W s_\beta t_\beta s_W^2} \{ (-1+t_\beta \kappa_d^A) C_{12} +
 t_\beta \kappa_d^A (2 C_2 + C_{22}) \}  \\
&&\hspace*{-3cm} g_L^{b H^0 H^\pm} = \frac{\alpha s_{\beta-\alpha} m_b \kappa_d^H}{16 M_W \pi s_W^2 t_\beta} \{m_b^2 t_\beta \kappa_d^A (-C_0 + C_{11} + C_{12} - C_2)
  + m_t^2 (C_{12} + C_2 + C_{22})\}  \\
&&\hspace*{-3cm} g_L^{A^0 H^\pm t} = \frac{-\alpha m_b m_t^2}{16 M_W \pi s_W^2 t_\beta^2}\{(1+t_\beta \kappa_d^A) C_{12} + t_\beta \kappa_d^A C_{22} \}  \\
&&\hspace*{-3cm} g_L^{A^0 b H^\pm} = \frac{\alpha m_b \kappa_d^A}{16 M_W \pi s_W^2 t_\beta} \{m_b^2 t_\beta \kappa_d^A C_{11} + (m_t^2 + m_b^2 t_\beta \kappa_d^A) C_{12}\} \\
&&\hspace*{-3cm} g_L^{h^0 t G^\pm} = \frac{-\alpha c_\alpha m_b m_t^2 s_{\beta-\alpha}}{16 M_W \pi s_\beta s_W^2} \{2 C_1 + C_{11} + 2 C_{12} \} \\
&&\hspace*{-3cm} g_L^{b h^0 G^\pm} = \frac{-\alpha m_b s_{\beta-\alpha} \kappa_d^h}{16 M_W \pi s_W^2} \{m_b^2 (C_0 - C_{11} - C_{12} + C_2) + m_t^2 (C_{12} +C_2+C_{22})\} \\
&&\hspace*{-3cm} g_L^{H^0 t G^\pm} = \frac{-\alpha s_\alpha m_b m_t^2 c_{\beta-\alpha}}{16 M_W \pi s_\beta s_W^2} (2 C_1 + C_{11} + C_{12}) \\
&&\hspace*{-3cm} g_L^{b H^0 G^\pm} = \frac{-\alpha m_b c_{\beta-\alpha} \kappa_d^H}{16 M_W \pi s_W^2} \{m_b^2 (C_0 - C_{11} - C_{12} + C_2) + m_t^2 (C_{12} +C_2+C_{22})\} \\
&&\hspace*{-3cm} g_L^{t G^\pm G^0} = \frac{\alpha m_b m_t^2}{16 M_W \pi s_W^2} \{C_0 + C_{11} + C_{22} + 2 (C_1 + C_{12} + C_2)\} 
 \eea
 \bea
&&\hspace*{-3cm} g_L^{b G^\pm G^0} = \frac{\alpha m_b}{16 M_W \pi s_W^2} \{m_b^2 (C_0 + C_1 + C_{12} + 2 C_2 + C_{22})+ m_t^2 (C_1 + C_{11} +C_{12})\} \\
&&\hspace*{-3cm} g_L^{\gamma b t} = \frac{Q_t Q_b \alpha m_b M_W}{2 \pi} \{C_1 + C_{11} + C_{12}\}\\
&&\hspace*{-3cm} g_L^{b t Z} = \frac{\alpha m_b M_W (-3 + 4 s_W^2)}{72 c_W^2 \pi s_W^2} \{ (-3+2s_W^2) (C_{12} +C_{22}) + 2 s_W^2 C_2\} \\
&&\hspace*{-3cm} g_L^{g b t} = \frac{- C_F \alpha_s m_b M_W}{2 \pi} \{C_1 + C_{12} + C_{11} \} \\
&&\hspace*{-3cm} g_L^{\gamma b G^\pm } = \frac{Q_b \alpha m_b M_W}{4 \pi} \{C_0 + C_1 +C_2\} \\
&&\hspace*{-3cm} g_L^{bG^\pm Z} = \frac{-\alpha m_b M_W s_W^2}{12 c_W^2 \pi} C_2 \\
&&\hspace*{-3cm} g_L^{h^0 t W} = g_L^{H^0 t W} = 0 \\
&&\hspace*{-3cm} g_L^{\gamma t G^\pm} = \frac{-\alpha Q_t m_b M_W}{4 \pi} \{C_0 + C_1 + C_2\} \\
&&\hspace*{-3cm}  g_L^{t G^\pm Z} = \frac{\alpha m_b M_W (-3 + 4 s_W^2)}{24 c_W^2 \pi} C_2 \\
&&\hspace*{-3cm} g_L^{b  h^0 W^\pm} = \frac{\alpha m_b M_W s_{\beta - \alpha} \kappa_d^h}{8 \pi s_W^2} C_2 \\
&&\hspace*{-3cm}  g_L^{b H^0 W^\pm} = \frac{\alpha  m_b M_W c_{\beta-\alpha} \kappa_d^H}{8 \pi s_W^2} C_2 \\
&&\hspace*{-3cm} g_L^{\gamma t W^\pm} = \frac{Q_t \alpha m_b M_W}{4 \pi} \{C_0 + C_1 - 2 C_{12} - C_2\} \\
&&\hspace*{-3cm} g_L^{\gamma b W^\pm} = \frac{Q_b \alpha m_b M_W}{4 \pi} \{-C_0 + C_1 + C_2 + 2 (C_{11} + C_{12}) \} \\
&&\hspace*{-3cm} g_L^{t W^\pm Z} = \frac{-\alpha m_b M_W (-3 + 4 s_W^2)}{24 \pi s_W^2} \{ 2 C_{11} + 2 C_{12} - C_2 \} \\
&&\hspace*{-3cm} g_L^{b Z W^\pm} = \frac{\alpha m_b M_W}{24 \pi s_W^2} \{-(3 + 4 s_W^2) C_1 + 2 (3 - 2 s_W^2) (C_{12}+C_{22}) - 6 s_W^2 C_2 \}
 \eea

\subsection{Tensorial Coupling $g_R$}
\vspace{-2cm}
\bea
 &&\hspace*{-3cm} g_R^{bh^0t} = \frac{\alpha c_\alpha m_b^2 m_t \kappa_d^h}{16 M_W \pi s_\beta s_W^2} \{ C_0 - C_{11} - C_{12} + C_2 \} \\
 &&\hspace*{-3cm} g_R^{bH^0t} = \frac{\alpha s_\alpha m_b^2 m_t \kappa_d^H}{16 M_W \pi s_\beta s_W^2} \{C_0 - C_{11}-  C_{12} + C_2 \}\\
 &&\hspace*{-3cm} g_R^{A^0 b t} = \frac{-\alpha m_b^2 m_t \kappa_d^A}{16 M_W \pi t_\beta s_W^2} \{ C_{11} + C_{12} \}\\
 &&\hspace*{-3cm} g_R^{b t G^0} = \frac{\alpha m_b^2 m_t}{16 M_W \pi s_W^2} \{C_0 + C_1 + C_{12} + 2 C_2 + C_{22} \}
\eea
\bea
&&\hspace*{-3cm} g_R^{h^0 H^\pm t} = \frac{\alpha c_\alpha c_{\beta-\alpha} m_t}{16 \pi M_W s_\beta t_\beta s_W^2} \{ (m_t^2-m_b^2 t_\beta \kappa_d^A) C_{12} +
 m_t^2 (2 C_2 + C_{22}) \}  \\
 &&\hspace*{-3cm} g_R^{b h^0 H^\pm} = \frac{-\alpha c_{\beta-\alpha} m_b \kappa_d^h}{16 M_W \pi s_W^2 t_\beta} \{ - C_0 + C_{11} +C_{12} - C_2 
 + t_\beta \kappa_d^A (C_{12} + C_2 + C_{22}) \} \\
 &&\hspace*{-3cm} g_R^{H^\pm H^0 t} =  \frac{-\alpha s_\alpha s_{\beta-\alpha} m_t}{16 \pi M_W s_\beta t_\beta s_W^2}  \{ (m_t^2-m_b^2 t_\beta \kappa_d^A) C_{12} +
 m_t^2 (2 C_2 + C_{22}) \} \\
 &&\hspace*{-3cm} g_R^{b H^0 H^\pm} = \frac{\alpha s_{\beta-\alpha} m_b^2 m_t \kappa_d^H}{16 M_W \pi s_W^2 t_\beta} \{- C_0 + C_{11} + C_{12} - C_2 + t_\beta \kappa_d^A (C_2 + C_{12}+ C_{22}) \}\\
 &&\hspace*{-3cm} g_R^{A^0 H^\pm t} = \frac{\alpha m_t}{16 M_W \pi s_W^2 t_\beta^2}\{ (m_t^2+  m_b^2 t_\beta \kappa_d^A) 
 C_{12} + m_t^2 C_{22} \} \\
&&\hspace*{-3cm} g_R^{A^0 b H^\pm} = \frac{-\alpha m_b^2 m_t \kappa_d^A}{16 M_W \pi s_W^2 t_\beta}
 \{C_{11} + (1+t_\beta \kappa_d^A)C_{12} \} \\
 &&\hspace*{-3cm} g_R^{h^0 t G^\pm} = \frac{\alpha c_\alpha m_t s_{\beta-\alpha}}{16 M_W \pi s_\beta s_W^2} \{m_t^2 (2 C_1 + C_{11} + C_{12}) + m_b^2 C_{12}  \} \\
 &&\hspace*{-3cm} g_R^{b h^0 G^\pm} = \frac{\alpha m_b^2 m_t s_{\beta-\alpha} \kappa_d^h}{16 M_W \pi s_W^2} \{C_0 - C_{11} + 2 C_2 + C_{22}\} \\
 &&\hspace*{-3cm} g_R^{H^0tG^\pm} = \frac{\alpha s_\alpha m_t c_{\beta-\alpha}}{16 M_W \pi s_\beta s_W^2} \{m_t^2 (2C_1 + C_{11} + C_{12}) + m_b^2 C_{12} \} \\
&&\hspace*{-3cm} g_R^{b H^0 G^\pm} = \frac{\alpha m_b^2 m_t c_{\beta-\alpha} \kappa_d^H}{16 M_W \pi s_W^2} 
 \{C_0 - C_{11} + 2C_2 + C_{22}\} \\
 &&\hspace*{-3cm} g_R^{b G^\pm G^0} = \frac{\alpha m_b^2 m_t}{16 M_W \pi s_W^2} \{ C_0 + 2 C_1 + C_{11} + 2 C_{12} + 2 C_2 + C_{22}\} \\
 &&\hspace*{-3cm} g_R^{\gamma b t} = \frac{Q_t Q_b \alpha m_t M_W}{2 \pi} \{C_2 + C_{22} + C_{12}\} \\
&&\hspace*{-3cm} g_R^{b t Z}  = \frac{-\alpha m_t M_W (-3 + 2 s_W^2)}{72 c_W^2 \pi s_W^2} \{ (-3+4s_W^2) C_{12}  - 3 C_2\} \\
&&\hspace*{-3cm} g_R^{g b t} = \frac{- C_F \alpha_s m_t M_W}{2 \pi} \{C_2 + C_{12} + C_{22} \} \\
&&\hspace*{-3cm} g_R^{\gamma b G^\pm} = \frac{- Q_b \alpha m_t M_W}{4 \pi} \{C_0 + C_1 +C_2\} \\
&&\hspace*{-3cm} g_R^{bG^\pm Z} = \frac{\alpha m_t M_W (-3 + 2 s_W^2)}{24 c_W^2 \pi} C_2 \\
&&\hspace*{-3cm} g_R^{h^0 t W} = \frac{\alpha c_\alpha m_t M_W s_{\beta-\alpha}}{8 \pi s_\beta s_W^2} C_2 \\
&&\hspace*{-3cm} g_R^{H^0 t W} = \frac{\alpha s_\alpha m_t M_W c_{\beta-\alpha}}{8 \pi s_\beta s_W^2} C_2 \\
&&\hspace*{-3cm} g_R^{\gamma t G^\pm} = \frac{\alpha Q_t m_t M_W}{4 \pi} \{C_0 + C_1 + C_2\} \\
&&\hspace*{-3cm} g_R^{t G^\pm Z} = \frac{-\alpha m_t M_W s_W^2}{6 c_W^2 \pi} C_2 \\
&&\hspace*{-3cm} g_R^{b h^0 W^\pm} = g_R^{b H^0 W^\pm} = 0 \\
&&\hspace*{-3cm} g_R^{\gamma t W^\pm} = \frac{Q_t \alpha m_t M_W}{4 \pi} \{-C_0 + C_1 + C_2 + 2 (C_{12} + C_{11}) \}
\eea
\bea
&&\hspace*{-0.75cm} g_R^{\gamma b W^\pm} = \frac{Q_b \alpha m_t M_W}{4 \pi} \{C_0 + C_1 - 2 C_{12} - C_2 \} \\
&&\hspace*{-0.75cm} g_R^{t W^\pm Z} = \frac{-\alpha m_t M_W}{24 \pi s_W^2} \{ (3 + 8 s_W^2) C_1 - 2 (3 - 4 s_W^2) (C_{12}+C_{22}) + 12 s_W^2 C_2 \} \\
&&\hspace*{-0.75cm} g_R^{b Z W^\pm} = \frac{\alpha m_t M_W (-3 + 2 s_W^2)}{24 \pi s_W^2} \{-2(C_{11}+C_{12}) +  C_2 \}
 \eea
\subsection{Right Chiral Coupling $V_R$}

\begin{eqnarray}
 V_R^{bh^0t} &=& \frac{-\alpha c_\alpha m_b m_t \kappa_d^h}{16 M_W^2 \pi
  s_\beta s_W^2} \{-B_0(m_b^2,m_b^2,m_{h^0}^2)+2C_{00} +  \nonumber\\ &&
m_b^2(C_1+C_{11}+C_{12}) + M_W^2(C_1+C_0+C_2) - m_t^2(C_1+C_0+C_{12})\}\nonumber \\
V_R^{bH^0t} &=& \frac{-\alpha s_\alpha m_b m_t \kappa_D^H}{16 M_W^2 \pi
  s_\beta s_W^2} \{-B_0(m_b^2,m_b^2,m_{h^0}^2)+2C_{00} +  \nonumber\\ &&
m_b^2(C_1+C_{11}+C_{12})+M_W^2(C_1+C_0+C_2) - 
m_t^2(C_1+C_0+C_{12})\}\nonumber \\
V_R^{A^0bt} &= &\frac{-\alpha m_b m_t \kappa_d^A}{16 M_W^2 \pi s_W^2 t_\beta}
\{-B_0(m_b^2,m_b^2,m_{A^0}^2) + 2C_{00} - M_W^2 C_1 \nonumber\\ &&
+ m_b^2 (C_1 - C_{11} - C_{12}) + 
m_t^2 (C_1 - C_{12} - 2 C_2 - C_{22} )\}\nonumber \\
V_R^{b tG^0} &=& \frac{-\alpha m_b m_t}{16 M_W^2 \pi s_W^2} \{B_0(m_b^2,m_b^2,M_Z^2)- 2C_{00} - m_b^2 (C_{12} +C_2 +C_{22})  + 
\nonumber\\ && m_t^2 (C_0 + C_{12} + C_2) - M_W^2 (C_0 + C_1 + C_2)\} \nonumber \\
V_R^{h^0H^\pm t} &=& \frac{\alpha c_\alpha c_{\beta-\alpha} m_b m_t}{16 M_W^2 \pi s_\beta s_W^2} 
\{ t_\beta \kappa_d^A (2 C_{00} + (m_t^2 - m_b^2) C_{12}) + 
 m_t^2 (1 + \kappa_d^A t_\beta) (2 C_2 + C_{22}) \} \nonumber\\
V_R^{bh^0H^\pm} &= &\frac{\alpha c_{\beta-\alpha} m_b m_t \kappa_d^h}{16 M_W^2 \pi t_\beta s_W^2} \{m_b^2 ((C_0 - C_{11} - C_{12} - C_2) 
+ t_\beta \kappa_d^A (C_0 - C_{11} - 2  C_{12} - C_{22}))  \nonumber \\
&& - m_t^2 (C_{12} + C_2 + C_{22}) \} \nonumber\\
V_R^{H^\pm H^0t} &=& \frac{-\alpha s_{\beta-\alpha} s_\alpha m_b m_t}{16 M_W^2 \pi t_\beta s_\beta s_W^2} \{2 t_\beta \kappa_d^A C_{00}
+ (m_t^2 - m_b^2) t_\beta \kappa_d^A C_{12} + m_t^2 (1 + t_\beta \kappa_d^A) (2 C_2 + C_{22})\} \nonumber\\
V_R^{A^0 b H^\pm} &=& \frac{\alpha m_b m_t \kappa_d^A}{16 M_W^2 \pi s_W^2 t_\beta} \{-2 C_{00} + m_b^2 (-1+\kappa_d^A t_\beta)C_{11}+ (m_t^2-m_b^2)C_{12}\}\nonumber \\
V_R^{h^0 t G^\pm} &=& \frac{\alpha c_\alpha m_b m_t s_{\beta-\alpha}}{16 M_W^2 \pi s_\beta s_W^2} \{-2C_{00} + (m_b^2-m_t^2)C_{12}\} \nonumber\\
V_R^{b h^0 G^\pm } &=& \frac{\alpha s_{\beta-\alpha} m_b m_t \kappa_d^h}{16 M_W^2 \pi s_W^2} \{-2C_{00} + (m_b^2-m_t^2)(C_{12}+C_2+C_{22})\}\nonumber
\end{eqnarray}
\begin{eqnarray}
V_R^{H^0 t G^\pm} &=& \frac{\alpha c_{\beta-\alpha} m_b m_t s_\alpha}{16 M_W^2 \pi s_W^2 s_\beta} \{-2C_{00} +(m_b^2-m_t^2)C_{12}\}\nonumber  \\
V_R^{b H^0 G^\pm} &=& \frac{\alpha c_{\beta-\alpha} m_b m_t \kappa_d^H }{16 M_W^2 \pi s_W^2} \{-2C_{00} + (m_b^2 -m_t^2)(C_{12}+C_2+C_{22})\}\nonumber \\
V_R^{t G^\pm G^0} &=& \frac{\alpha m_b m_t}{16 M_W^2 \pi s_W^2} \{2 C_{00} + m_b^2 (C_1 + C_{11} + C_{12}) + m_t^2 (2 C_0 + 3 C_1 + C_{11} + 3 C_{12} + 4 C_2 + 2 C_{22})\} \nonumber\\
V_R^{b G^0 G^\pm} &=& \frac{\alpha m_b m_t}{16 M_W^2 \pi s_W^2} \{2 C_{00} +
m_b^2 (2 C_0 + 3 C_1 + C_{11} + 3 C_{12} + 4 C_2 + 2 C_{22}) + m_t^2 (C_1 +C_{11} + C_{12})\}  \nonumber\\
V_R^{\gamma b t} &=& \frac{Q_t Q_b \alpha m_b m_t}{2 \pi} \{C_{11} + 2 C_{12}+C_{22} \} \nonumber\\
V_R^{b t Z} &=& \frac{\alpha m_b m_t}{72 \pi c_W^2 s_W^2} \{8 s_W^4 C_0 +2 s_W^2 (-9 + 8 s_W^2) C_2 + (9  - 18 s_W^2  + 8 s_W^4) C_{22}\}\nonumber \\
V_R^{g b t} &=& \frac{- C_F \alpha_s m_b m_t}{2 \pi} \{C_{22} + 2 C_{12} + C_{11} \} \nonumber\\
V_R^{b G^\pm \gamma} &=& \frac{Q_b m_b m_t}{4 \pi} C_1\nonumber \\
V_R^{b G^\pm Z} &=& \frac{-\alpha m_b m_t s_W^2}{12 c_W^2 \pi} \{C_0+C_1 + C_2\}\nonumber \\
V_R^{t h^0 W^\pm} &= & V_R^{t H^0 W^\pm} =  0\nonumber \\
V_R^{\gamma t G^\pm} &=& \frac{Q_t \alpha m_b m_t}{4 \pi} C_1\nonumber \\
V_R^{t G^\pm Z} &=& \frac{-\alpha m_b m_t s_W^2}{6 c_W^2 \pi} \{C_0+C_1 +  C_2\}\nonumber \\
V_R^{b h^0 W^\pm} &=& V_R^{b H^0 W^\pm} = 0\nonumber \\
V_R^{\gamma t W^\pm} &=& \frac{- Q_t \alpha m_b m_t}{4 \pi} (C_1 - 2 C_{11})\nonumber \\
V_R^{\gamma b W^\pm} &=& \frac{-Q_b \alpha m_b m_t }{4 \pi} \{C_1 - 2 C_{11}\}\nonumber \\
V_R^{t W^\pm Z} &=& \frac{\alpha m_b m_t}{12 \pi s_W^2} \{-6 s_W^2 C_0 + (3 -
 4 s_W^2) (C_{22} + C_{11} + 2 C_{12})  + (3 - 10 s_W^2) (C_1 + C_2))\}\nonumber \\
V_R^{b W^\pm Z} &=& \frac{\alpha m_b m_t}{12 \pi s_W^2} \{-3 s_W^2 C_0 + (3 -5 s_W^2) (C_1+C_2) + (3-2 s_W^2) (C_{11} + 2 C_{12} + C_{22}) \}\nonumber
\end{eqnarray}

Where $C_{i, ij} = C_{i, ij} (m_b^2, m_t^2, M_W^2, m_A^2, m_B^2, m_C^2)$, $A, B \textrm{ and } C$ are the particles running in the loops.
$Q_b=1/3, Q_t=2/3 \text{ and } C_F=4/3$,
$s_W =\sin \theta_W, c_W = \cos \theta_W, c_i=\cos i, s_i = \sin i \textrm{ and } t_i = \tan i \textrm{ where } i=\alpha,\beta$. 
Expressions in the case of the Standard Model are recovered by letting $s_{\beta-\alpha}\to1, c_{\beta-\alpha}\to0$
and $s_{\beta}=c_{\alpha}, s_{\alpha}=-c_{\beta}$ in the previous formulae.

\end{document}